# Compounding Injustice

**History and Prediction in Carceral Decision-Making**

Benjamin David Laufer

April 16, 2019

Submitted in partial fulfillment
of the requirements for the degree of
Bachelor of Science in Engineering
Department of Operations Research and Financial Engineering
Princeton University

*I hereby declare that I am the sole author of this thesis. I authorize Princeton University to lend this thesis to other institutions or individuals for the purpose of scholarly research.*

---

Benjamin Laufer

*I further authorize Princeton University to reproduce this thesis by photocopying or by other means, in total or in part, at the request of other institutions or individuals for the purpose of scholarly research.*

---

Benjamin Laufer

# Acknowledgments

First, I want to thank my family. Mom, Dad, Jacob, and William - Thanks for showing your love at every moment over the past twenty-two years.

I'd like to thank my advisor, Miklos Racz. From the beginning, you welcomed and encouraged creativity. Our frequent meetings became a sounding board for my numerous (and at times erratic) ideas. I admire your ability to listen to concepts and, at face value, carefully parse out their merit. You've offered a totally unnecessary level of support, for which I'm deeply grateful.

I also must thank Eduardo Morales for his enthusiasm and warmth. He encountered my thesis exactly three weeks before its due date and somehow found the time and bandwidth to meet with me repeatedly and help me whip it into shape.

Thank you to all my friends at Princeton and Hunter. A few names of people who have helped out, in no particular order: Danny Li, Amanda Brown, Micah Herskind, Maya Von Ziegesar, Nate Moses, Louis Aaron and Marie-Rose Sheinerman.

I am incredibly grateful to have been able to attend talks or correspond directly with people who know a lot about the topic of my thesis. In particular, I'd like to thank: Annette Zimmermann, Arvind Narayanan, Jessica Eaglin, and Jon Kleinberg.

Finally, I want to express gratitude towards the Petey Greene Program, and everybody I've met because of my involvement. Jim Farrin, having founded Petey Greene eleven years ago at Princeton, built a national network of students who devote hours each week to teach, and learn from, incarcerated folks. I'm grateful to the men I've worked with over the past four years who have profoundly influenced by views on this topic.

# Forward

When I set out to work on this thesis, I planned to research the trade-off between algorithmic accuracy and transparency in criminal justice. I assumed that the criminal system was caught between accurately predicting crime, on the one hand, and requiring algorithms to be simple, explainable, and unbiased on the other. After significant research, I ended up deeply questioning my own beliefs about criminality, criminal justice and prisons. Ultimately, I found that one concept I had assumed to be a central tenet in good criminal treatment - *accurate prediction* - might be utterly unsalvageable for criminal policy.

Thus, many of the coming pages are critical, and few are typical for an engineering thesis. My goal is to leave a bit of what I've learned for anyone who might read this. I also hope to demonstrate that a rigorous, quantitative education can and should remain socially and politically engaged. We know how to solve problems, so we should also know how to identify them.

Finally, I want to note that many of the topics I bring up just barely scratch at the surface of our deeply complex system of criminal punishment. I hope to always continue discussing, writing and learning in the future.

# Abstract

Risk assessment algorithms in criminal justice put people's lives at the discretion of a simple statistical tool. This thesis explores how algorithmic decision-making in criminal policy can exhibit feedback effects, where disadvantage accumulates among those deemed 'high risk' by the state. Evidence from Philadelphia suggests that risk – and, by extension, criminality – is not fundamental or in any way exogenous to political decision-making. A close look at the geographical and demographic properties of risk calls into question the current practice of prediction in criminal policy. Using court docket summaries from Philadelphia, we find evidence of a criminogenic effect of incarceration, even controlling for existing determinants of 'criminal risk'. With evidence that criminal treatment can influence future criminal convictions, we explore the theoretical implications of compounding effects in repeated carceral decisions.

# HOLMESBURG, PHILADELPHIA

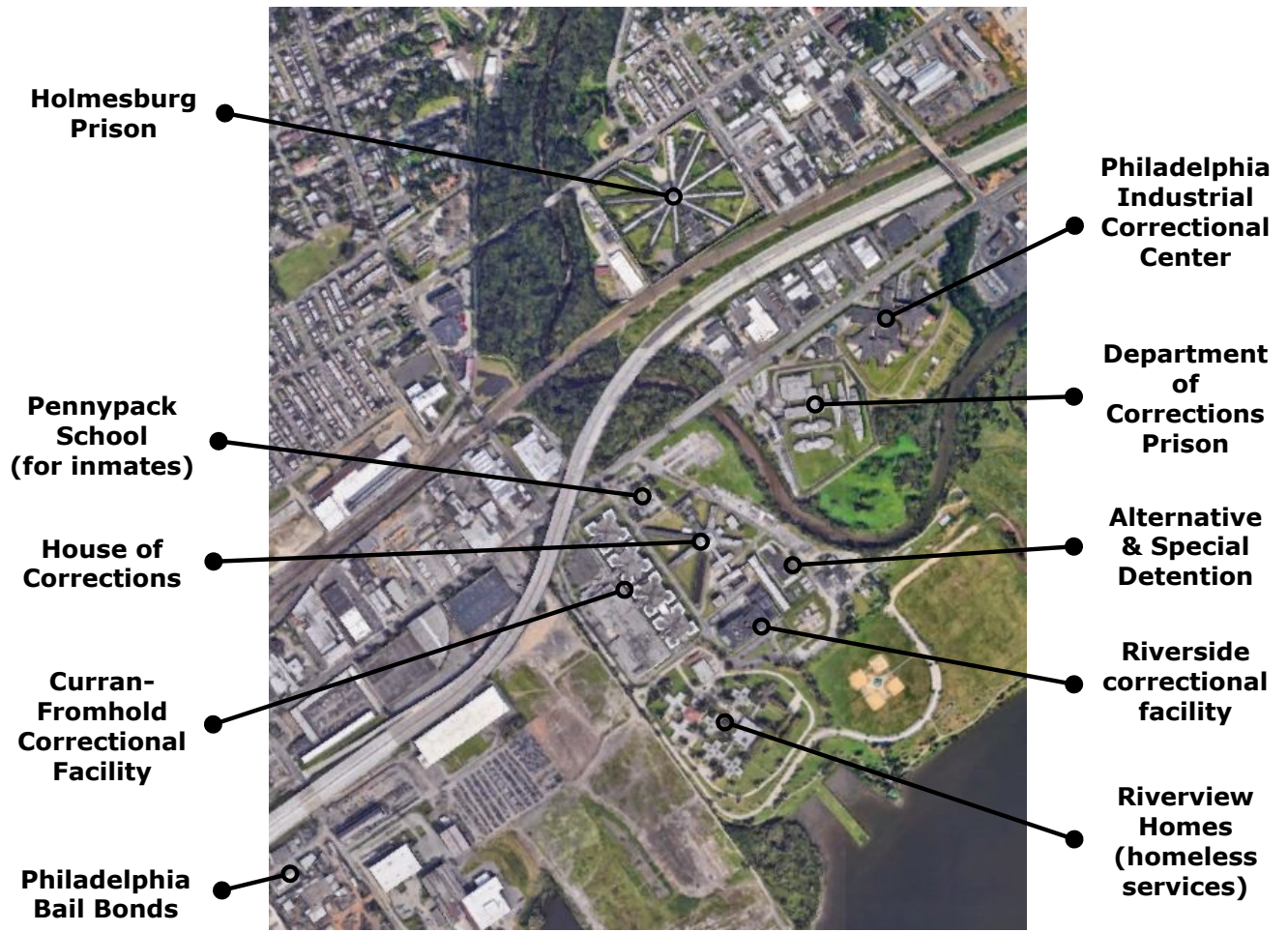

- Holmesburg Prison
- Philadelphia Industrial Correctional Center
- Pennypack School (for inmates)
- Department of Corrections Prison
- House of Corrections
- Alternative & Special Detention
- Curran-Fromhold Correctional Facility
- Riverside correctional facility
- Riverview Homes (homeless services)
- Philadelphia Bail Bonds

# Contents

# Part I

# Predictions

Algorithmic Risk Assessment in the Status Quo

# Chapter 1

# Historical Context

William Horton was imprisoned in Massachusetts for first degree murder. Released for a weekend on furlough in 1986, Horton failed to return and was ultimately re-arrested after raping a white woman and assaulting her fiancé (Newburn & Jones, 2005). In 1988, campaigners for presidential nominee George H.W. Bush broadcast a now notorious TV ad featuring Horton. By including Horton's mugshot and giving him the nickname 'Willie,' the ad managed to play off racial fears without explicitly mentioning Horton's race.[1] Bush's campaign used the story to criticize the Democratic nominee, Michael Dukakis, who was governor of Massachusetts and had helped to craft the furlough program that enabled Horton's misconduct. Despite the fact that most states allowed the same program (Ried, 1988), and the exceedingly low chances of behavior like Horton's, his name was a hit on the campaign trail and his behavior became a defining issue in the election. Bush's lead political strategist, Lee

---

[1] The ad is notorious for defining 'dog-whistle' politics. It did not explicitly reference William Horton's race, but race was undoubtedly relevant to its reception. See (Anderson & Enberg, 1995) and (Hurwitz & Peffley, 2005).

Atwater, famously said, "By the time we're finished, they're going to wonder whether Willie Horton is Dukakis's running mate" (Schwartzapfel & Keller, 2015). After being elected, President Bush in 1989 called for "more jails, more prisons, more courts and more prosecutors" (Dvorak, 2018), setting the stage for the decades of tough-on-crime policies that would follow.

Criminality and criminal punishment are messy concepts - they depend heavily on political and social factors that change over time. Judges, having broad authority over decisions about sentencing, parole, pretrial detention and bail, are guided by a number of goals - public safety, humane treatment, retribution, rehabilitation, deterrence, and others. As the various and often-conflicting aims of criminal punishment come in and out of vogue, laws follow. And, as algorithms encroach on an increasing number of human decisions, engineers encounter the same, entangled mess of conflicting values and historical complexities that judges have to reconcile every day. When each modeling decision might determine somebody's freedom, the use of algorithms in the criminal justice system is necessarily political, and inseparable from social and historical context.

Algorithmic risk assessment in criminal policy has a surprisingly long history. The departure from a system reliant solely on professional expertise began in 1928, when Ernest Burgess developed a tool to predict recidivism in order to aid parole decisions in Nebraska (Burgess, 1928). A professor of sociology at the University of Chicago, Burgess used statistical regression techniques to predict violations of parole (Harcourt, 2014). Students of Burgess and other sociologists began to update and improve actuarial risk assessment methods, and ultimately developed similar methods for other judicial decisions like sentencing and bail.

Following Burgess's lead, the actuarial methods developed in the following decades used race as an explicit determinant for treatment. Burgess was obsessively fixated on the race and national origin of parolees, and how these factors correlated with recidivism. He found black people and immigrants to be the two most likely groups to recidivate, and his tool's adoption in Nebraska used his findings to decline parole for members of these groups (Harcourt, 2014, 238). The tools developed thereafter all used information about familial nationality and ethnic origin - even as late as the 1970s, California developed its "Base/Expectancy Score," which used race as one of only four factors to determine criminal risk (Harcourt, 2014, 238). There are numerous ways to explain why these algorithms included race - sociologists could have been driven by hateful views of blacks, they could have believed in genetically determinant theories of criminal behavior, they could have inherited understandings of racial difference that were widely accepted at the time, or all of the above. But, regardless of the personal beliefs of individuals, these algorithms were adopted because they *did* find statistically significant relationships between race and crime. In the new, evidence-based generation of risk assessment, theoretical links were deemed unimportant compared to the simple statistical accuracy of predictive tools (Bonta & Andrews, 2007, 4). By focusing only on statistical performance, developers of risk tools encoded and formalized discrimination in a new way.

As theorists and judges began to identify limitations in using demographics and immutable ('static') characteristics for assessing risk, a 'third generation'[2] of risk assessment tools ushered changes in criminal treatment in the late 1970s and early 1980s. These algorithms, no longer using race as an input, found that the individual

---

[2] The framing of risk assessments in four 'generations' is used widely in the literature. The first generation was characterized by widespread reliance on professional judgment, and the second generation came with the introduction of actuarial, evidence-based methods. See (Bonta & Andrews, 2007).

*criminogenic needs* of a defendant were important in addition to underlying risk.[3] Thus, the factors used in actuarial decisions were called 'risk-needs factors'[4] and included information that could change over time - employment status, family status, peer networks, substance abuse, psychological conditions, and residential location (Picard-Fritsche et al., 2017). While embracing the possibility that risk can decrease over time, criminogenic needs are notable because they embed dynamic social designations within the assessment of risk. The added factors, from employment to mental health, are themselves reliant on a host of decisions and designations by employers psychiatrists, landlords, and others. The adoption and evolution of statistical risk assessment therefore did not eradicate human decisions (and their accompanying biases) from criminal treatment. Quite the opposite; algorithms helped judges synthesize a whole host of labels that can be socially and historically fraught.

The 1980s saw a shift in public opinion that resulted in tough-on-crime policies. By this time, the principle of risk was deeply embedded in the public's understanding of crime and, as a result, the institutional treatment of criminals. The dynamics were apparent in the newly popular theory of "Selective Incapacitation" - coined by criminologists Peter Greenwood and Allan Abrahamse in 1982, the theory discredits rehabilitation catered to criminogenic needs (Greenwood, Abrahamse, et al., 1982, vii-viii), and instead emphasizes identifying the highest-risk offenders *early* to minimize harm to the community. The authors write:

> Selective incapacitation is a strategy that attempts to use objective actuarial evidence to improve the ability of the current system to identify and confine offenders who represent the most serious risk to the community.(Greenwood et al., 1982, vii)

---

[3] 'Criminogenic effects' broadly refer to effects that cause crime. 'Criminogenic needs' therefore refers to aspects of a defendant that may lead to further criminal behavior, without proper intervention. See, for example, (Wooditch, Tang, & Taxman, 2014).

[4] See (Picard-Fritsche, Tallon, Adler, & Reyes, 2017, 5-6).

Where questions of parole and furlough deal with the treatment of defendants who are already incarcerated, the theory of selective incapacitation is remarkable because it advocates crafting sentences (i.e., punishments) from a mere *anticipation* of future crime. Like the risk tools in use then and those in use today, incapacitation theory boasts that it is 'objective,' 'actuarial,' and based on 'evidence.' It downplays its own harms by marketing itself as simply an improvement to existing processes. And, by situating former offenders as a 'serious risk to the community,' it implies that these people are external to the community and represent a threat that (albeit predictable) is outside of our control.

Tough-on-crime politics and legislation in the 1980s and 1990s dramatically increased the prison population and disproportionately impacted black communities. Whereas in the 1930s, black Americans were three times more likely to be incarcerated than whites, in the 1990s they were seven times more likely (Lyons & Pettit, 2011, 258). Tough on crime politics were fueled by (and helped fuel) racial stereotypes (Hurwitz & Peffley, 1997). And, to this day, black people are more likely to be searched (Engel & Johnson, 2006), arrested (ACLU, 2015), detained (Arnold, Dobbie, & Yang, 2018), and incarcerated. The public's fear of crime was inextricably linked to race, demonstrated by the success of the Willie Horton ad.

In 1990, another reform came to risk assessment - the principle of 'responsivity' was introduced, and virtually every tool developed since them has described itself as catering to the three principles of risk, needs, and responsivity. Coined by Andrews et al. (Andrews et al., 1990), responsivity aims to identify treatments that are more- or less-conducive to an individual defendant. Now, instead of just understanding a defendant's various risk factors for violating the law, algorithms can identify which defendants may be able to markedly improve as the result of certain specialized treatments. Importantly, however, the converse is also true: responsivity modeling may enable courts to identify defendants who are *beyond hope* in some sense, and recommend confinement. In other words, responsivity is compatible with and even complimentary to selective incapacitation, and risk-driven decision making.

The development and refinement of criminal risk - as a concept and as a tool - is innately tied to historical decisions made by theorists, practitioners, judges and defendants. When a jurisdiction begins sentencing using a risk assessment tool, the algorithm's adoption is contextualized by a long history of judicial decisions that may not have been 'objective' or unbiased. The very notion of criminality is contingent on historical decisions made by lawmakers and enforcement officers. As algorithms have evolved, criminal risk has begun to represent a growing group of designations and measurements that are not as objective as they may seem: psychiatric disorders, social characterizations, family structure, living arrangements, employment. None of these labels are free from historical contingency, bias, and inaccuracy. So when the state begins to design punishments based on these labels, we need to ask where they come from, and how they may change over time. When an algorithm evolves from using *ethnicity of father*[5] to using *criminal history of father*,[6] we need to carefully consider how biases may be encoded in seemingly-innocuous pieces of information. And we need to rigorously scrutinize claims of objectivity and predictive accuracy in the context of criminal punishment.

---

[5] Burgess's tools used explicit racial measurements by asking about the ethnicity and national origin of the defendant's father. See (Harcourt, 2014).
[6] COMPAS risk assessment tool has a section devoted to 'Family Criminality' which asks questions about history of familial arrests and convictions.

# Chapter 2

# Framing in the Literature

## *Machine Learning and Prediction*

The advent and preliminary success of machine learning models has led to significant speculation and excitement about applications. These algorithms are well-suited to problems that require prediction of an unknown. In medicine, predictions may enable treatments that are more equipped to help individual patients. In finance, predicting default can enable more profitable lending for banks. In education, predicting student success can inform decisions in curricula, teaching, admissions and more. The extensive set of unsolved problems in society naturally drives our excitement about innovation. Accordingly, our impulse is to consider some of the most serious and impactful applications first.

Theorists, policy makers and practitioners have adopted the view that predictive algorithms can aid vital public institutions that persistently under-perform. Jon

Kleinberg, a leading researcher in machine learning, co-authored a letter (Kleinberg, Ludwig, Mullainathan, & Obermeyer, 2015) that coins the term 'prediction policy problem'. The paper emphasizes that quantitative policy research has focused too heavily on questions of causal inference, and that predictive algorithms can answer many important questions in academia and policy. To illustrate what predictive policy decisions are, Kleinberg et al. use the very simple example of the decision to carry an umbrella to work. In such a case, the person does not care what actions *cause* rain, and instead is only interested in *predicting* rain. Their claim is that many decision-making problems may benefit from predictive models that do not necessitate a causal understanding. Their call for research includes applications in medical testing, education, violence prevention, lending practices, and criminal decisions including bail.

Others have similarly encouraged applying available statistical techniques to highly impactful social and political decisions. Jung et al. in a 2017 paper entitled "Simple Rules for Complex Decisions" acknowledge that human decisions can be sub-optimal, and advocate a method they call 'select-regress-and-round' (Jung, Concannon, Shroff, Goel, & Goldstein, 2017). They use the example of bail decisions to argue that regression methods can greatly improve human decisions. They find that, despite being simplistic, regression methods are robust to a wide variety of complex decisions; the implication being that regression methods can often improve decisions even when they do not precisely model variable relationships.

Practitioners who develop algorithms for criminal justice have naturally joined the ML excitement. In a practitioner's guide for COMPAS, a widely used algorithm developed by Northpointe, Inc.[1], the company tries to describe the wide embrace of

---

[1] The company now operates under a parent company called Equivant. In this paper, we solely refer to the company as Northpointe, to be consistent with current literature on the topic.

data-driven decision-making:

> Statistically based risk/needs assessments have become accepted as established and valid methods for organizing much of the critical information relevant for managing offenders in correctional settings. Many researchers have concluded that objective statistical assessments are, in fact, superior to human judgment. COMPAS is a statistically based risk assessment developed to assess many of the key risk and needs factors in adult correctional populations and to provide information to guide placement decisions. (Northpointe, 2015, 1-2)

Sweeping statements about algorithms outperforming human predictions is cited as evidence for the adoption of algorithmic risk assessments. Similarly, emphasis on big data and academic research is used to establish the reputation of risk assessment tools. The Public Safety Assessment (PSA) boasts its data-driven development: "Researchers designed the PSA based on the largest, most diverse set of pretrial records ever assembled—750,000 cases from nearly 300 jurisdictions. Based on a comprehensive analysis of the data, researchers identified the nine factors that best predict pretrial risk" (Laura and John Arnold Foundation, 2019).

Evidently, people are excited about using algorithms in high-impact fields. But the question remains: are these algorithms being developed and adopted ethically? Innovations in highly impactful domains are promising, but also can cut dangerous corners, or have harmful unintended consequences. When Kleinberg et. al. leap from a simple example about umbrellas to some of the country's most complex decisions (in education, prison, medicine) we see a need to reconcile the complexity of the problem with the narrow scope of a predictive algorithm. Jung et al. actually advocate a simpler-than-realistic model, solely because the algorithm's performance beats humans in predictive accuracy. To deal with a highly complex and impactful problem like bail or sentencing, researchers are finding themselves needing to simplify and narrow their focus. People are fitting the problem to the model, instead of fitting a model to the problem.

## *Needs, Responsivity and Clinical Treatment*

Departing from the predictive view of algorithmic risk assessment, a few authors have begun to emphasize causal treatment effects in criminal interventions. In an article entitled "Beyond Prediction: Big Data for Policy Problems", Susan Athey notes that some policy problems must be concerned with values beyond prediction:

> It is sometimes important for stakeholders to understand the reason that a decision has been made . . . Transparency and interpretability considerations might lead analysts to sacrifice predictive power in favor of simplicity of a model. Another consideration is fairness, or discrimination. Consumer protection laws for lending in the United States prohibit practices that discriminate on the basis of race. Firms might wish to use SML methods to select among job applicants for interviews; but they might wish to incorporate diversity objectives in the algorithm, or at least prevent inequities by gender or race.

Athey's words draw attention to the complexity of policy problems. Goals may not align with one straight-forward objective and may require hard trade-offs. In a similar vein, Kleinberg et. al.'s "Algorithmic Fairness" offers a sort-of reconciliatory approach, claiming that algorithmic prediction is still important to these problems, but that other objectives can be introduced as constraints after-the-fact: "a preference for fairness should not change the choice of estimator. Equity preferences can change how the estimated prediction function is used (such as setting a different threshold for different groups) but the estimated prediction function itself should not change" (Kleinberg, Ludwig, Mullainathan, & Rambachan, 2018, 22-23) This theory is manifested in Kleinberg's work in algorithmic bail reform (Kleinberg, Lakkaraju, Leskovec, Ludwig, & Mullainathan, 2017).

Barbaras et al.'s "Interventions over Predictions: Reframing the Ethical Debate for Actuarial Risk Assessments" challenges the conventional treatment of risk assessment as a predictive policy problem (Barabas, Dinakar, Ito, Virza, & Zittrain, 2017).

The paper instead draws attention to the treatment effects of criminal interventions: "If machine learning is operationalized merely in the service of predicting individual future crime, then it becomes difficult to break cycles of criminalization that are driven by the iatrogenic effects of the criminal justice system itself" (Barabas et al., 2017, 1). Their central claim is that treating risk-assessment as a prediction policy problem does not actually answer the question of how we may be able to *lower* risk in the future: "Predictive risk assessments offer little guidance on how to effectively intervene to lower risk" (Barabas et al., 2017, 10). Barbaras et al. point out an important issue in Kleinberg and others' assumptions - The attempt to fit risk assessment into *prediction* necessarily removes criminogenic and institutional effects of criminal policy from individual behavior. Treating crime as an exogenous factor that is predictable-but-uncontrollable leads to a logic of incapacitation and abstention from any constructive or supportive interventions.

While identifying an important flaw in solely using predictive algorithms in criminal policy, clinical treatment of criminal policy is not a new idea. In the late 1970s and early 1980s, researchers began to realize that existing actuarial risk-assessment methods used only static, historical factors about defendants, and did not account for changes in behavior. After a generation of assessments that solely focused on the 'risk principle', algorithms in the 1980s began including information about 'needs' as well- dynamic information that might be linked to criminal behavior. These factors, summarized into seven 'major risk/need factors', are listed below.[2]

<div style="text-align:center">

Antisocial personality pattern
Procriminal Attitudes
Social Supports for Crime
Substance Abuse
Family/marital relationships

</div>

---

[2] See (Bonta & Andrews, 2007, 6) for a complete descriptions of these factors and their role in current criminal treatment.

School/work
Prosocial recreational activities

Here, we see very similar logic to Barbaras et al., that clinical treatment is necessary to cater interventions to individuals based on their propensities to commit crimes. Notice that this logic is not necessarily *incompatible* with the risk principal. What Bonta et al. describe as "Generation Four" algorithms purport to combine risk, needs, and responsivity to cater ideal criminal treatment to defendants.

Theories that emphasize clinical treatment over predictions still rely on *outcomes* to determine interventions. In Barbaras et al.'s paper, the writers advocate what they see as an alternative to predictive algorithmic risk assessment: "Rather than using machine learning for prediction, these methods could be used to identify features that are highly predictive of recidivism, in order to inform hypotheses on interventions (and their timing) that can then be tested using causal inference." The distinction between what they're critiquing and what they're defending is quite subtle - instead of "using machine learning for prediction", they advocate using the same methods to "identify features that are highly predictive of recidivism" to inform interventions. Indeed, treatment methods share assumptions with theories of selective incapacitation that trace back to the 1980s - that criminal justice policy should anticipate future crimes, and act in a way that protects society from potentially dangerous people. These methods assume that outcome variables - namely, recidivism - are objectively and equitably distributed. They assume that the designation of criminal action itself is in some way fundamental, and that police officers, juries, judges, prison guards and parole officers do not influence the *labelling* of people as criminals in problematic ways. We call these assumptions into question.

## *Bias in Criminal Policy*

Given the history of bias in criminal justice, many ethical questions have been brought to the field of risk assessments. A proposed sentencing tool in PA provoked public outcry recently because their new sentencing tool proposed using neighborhood as a covariate to assess risk.[3] Many have argued that other features act as a proxy for race, including criminal history.[4] People have also spoken against the proprietary nature of certain algorithms developed by private companies, including Northpointe's COMPAS. In perhaps the most high-profile and contentious critique of a risk-assessment algorithm, ProPublica released a report in 2016 entitled "Machine Bias: There's software used across the country to predict future criminals. And it's biased against blacks" (Angwin, Larson, Mattu, & Kirchner, 2016). In it, the authors analyzed COMPAS's false positive and false negative classifications among black and white defendants. They found that black defendants were more likely to be mislabeled as high-risk, while white defendants were more likely to be mislabeled as low risk.

Developers of risk assessment algorithms have defended their tools as objective and unbiased. In a scathing retort to Angwin et al.'s ProPublica article, Northpointe Inc. published a report that defended the COMPAS tool, and used the same data as the ProPublica article to establish 'accuracy equity' and 'statistical parity' (Dieterich, Mendoza, & Brennan, 2016).

Broadly, developers of risk assessments view these tools as helping to fix the problems of judicial bias and inconsistency, and often frame risk assessments as just another source of information for judges. Northpointe describes their COMPAS algorithm for case supervision review as "an objective decision support tool to guide adjustments in the current supervision level" (Northpointe, 2012).The Laura and John Arnold Foundation describe the Public Safety Assessment as a way to solve biases in the system: "Advocacy groups are raising important questions regarding potential racial bias and racial disparities in the use of risk assessments. It is within this broader context that pretrial risk assessment can play an important role" (Laura and John Arnold Foundation, 2019). Finally, Level of Service-Revised (LSI-R) produces a brochure to sell their algorithms to courts, and in it markets their tool's versatility: "Accurately assess any population: Valid and reliable in different countries, states, provinces, offender populations, genders, various minority groups, and settings" (MHS Public Safety, 2019). Words like "objectivity", "bias", "accuracy", "valid", and "reliable" are used generously by these sites, and for good reason - an algorithm that may be biased would not be adopted by courts. But we're left wondering what these words mean - in a system that has not seemed to rid itself of unequal treatment across racial and social lines, what do proponents of algorithms mean when they call a tool objective or valid?

---

[3] See (Christin, Rosenblat, & Boyd, 2015, 3). See also (Melamed, 2018).
[4] See (Harcourt, 2014).

Central to this question is the definition of fairness and fairness in algorithmic decision-making. In the ProPublica-Northpointe debate, theorists began to realize that the two organizations were operating with different definitions of bias - calibration, or 'predictive parity', ensures that defendants with the same score will have the same reoffense rates regardless of race. More specifically, if we model a defendant's recidivism outcome as a binary variable $Y \in \{1, 0\}$, and describe a defendant $d$ using risk score $s(d)$ and group membership (i.e. race) $g(d)$, then calibration is defined as:[5]

---

[5] See (Corbett-Davies, Pierson, Feller, Goel, & Huq, 2017, 798).

$$P(Y = 1|s(d), g(d)) = P(Y = 1|s(d))$$

Analyzing the debate between ProPublica and Northpointe Inc. on COMPAS, Kleinberg et al. found that the two organizations were using different definitions of fairness, and had stumbled upon a fundamental trade-off in algorithmic fairness. He finds that while Northpointe had been demonstrating their tool's calibration to argue that the scoring was unbiased, Angwin et al. at ProPublica had been analyzing false positive and false negative rates. He finds that, except in degenerate cases, an algorithm cannot guarantee the following three properties:[6]

- Calibration
- Balance for the negative class, meaning that among defendants who don't recidivate upon release, the average score is equal across groups.
- Balance for the positive class, meaning that among defendants who do recidivate upon release, the average score is equal across groups.

This finding has motivated research on inherent trade-offs in quantitative notions of bias. However, neither of these formal definitions fully shield algorithms from discriminative behavior. Indeed, Kleinberg et al. find that the only way to guarantee all conditions (A), (B) and (C) above, is to have "perfect prediction." Why? Say our scores $s(d)$ are binary and $s(d) = 1$ corresponds to $Y = 1$ with probability 1. In other words, $Y = s(d)$. Then our scores would of course be calibrated, since $P(Y = 1|s(d) = 1) = 1$ and $P(Y = 1|s(d) = 0) = 0$ in all cases, and the group designation $g(d)$ adds no information, since the score perfectly predicts outcome. Balance for the positive and negative classes is obvious - the average score for all defendants in the positive class is 1, and the average score for all defendants in the negative class is 0, regardless of race.

---

[6] Kleinberg et al. in (Kleinberg, Mullainathan, & Raghavan, 2016) is largely credited with finding this inherent conflict. However, the logic in (Corbett-Davies et al., 2017) is easier to follow.

Now suppose that the outcome variable $Y$ is somehow a *function* of race group g(d):

$$Y = f(g(d))$$

Our assumption here is that something about the label $Y$ is racially mediated. For example, if police officers are more likely to search a black man, then black men will be more likely to be convicted of crimes generally, and therefore recidivism $Y$ would depend on group designation. Let's start with a trivial case where black offenders will *definitely* recidivate, and other offenders will *never* recidivate:

$$Y = \mathbf{1}\{g(d) = black\}$$

In this case, a perfectly predictive algorithm will successfully pass both calibration and balance tests. Even if the algorithm uses information about peer networks, psychological and social analysis, familial crime history, and financial information - rather than race explicitly - to develop a perfect racial classifier, there's still something wrong with this scenario. Theorists have described the shortcomings of formal definitions of fairness, and the possibilities that variables can be 'reconstructed' through proxies.[7] In a similar vein, others have explored the idea that even perfectly unbiased predictive algorithms, as long as they treat groups differently, may be untenable.[8]

The formal notions of fairness in algorithmic decision-making highlights a short-coming in the literature: mathematical notions of fairness are seen as totally distinct and irrelevant to systemic and historically-contextual notions of unfairness in political conversations about criminal treatment. Recent worries that risk-assessment algorithms could encode historical judicial biases are legitimate, and represent a new challenge to tools like COMPAS that is distinct from the Angwin et al. critique. That is: not only do predictive inaccuracies challenge the efficacy of COMPAS, but a far-from-perfect system of labels in the criminal, medical and economic fields question the objectivity of any data-driven, risk-driven tool for criminal treatment decisions.

---

[7] An example in public policy and finance is red-lining - using location data as a proxy for race. See (Berkovec, Canner, Gabriel, & Hannan, 1994) and (Lang & Nakamura, 1993), for example. Of more relevance to the group of theorists we engage with in this work, see (Lakkaraju, Kleinberg, Leskovec, Ludwig, & Mullainathan, 2017).
[8] Information comes from correspondences with Annette Zimmermann, who has forthcoming work on this topic. See also (Li, 2019)

## *Validation and Instantial Experiments*

Risk assessment algorithms are developed and then tested for 'validity'. These experiments, formerly only concerned with predictive validity, now test various potential biases that algorithms may exhibit in new populations. Validation experiments have therefore become an important aspect of the risk-assessment development process, and validity is seen as a necessary requisite for any risk assessment algorithm in use. What does validity mean?

While there has been some controversy over the way in which risk assessment tools get developed,[9] remarkably little analysis has been conducted of the best practices for validation in risk assessment. As a result, many validation experiments resemble one another. Typically, the studies measure a tool's predictive capacity by analyzing post-conviction arrest rates over a short time-frame. They take a group of defendants released from the same jurisdiction in a given time-frame, and determine the average re-arrest rate of defendants with different risk scores over a typical period of one or two years. For example, Lowenkamp et al. conducted a validation experiment in which they tested the LSI-R and the LSI-Screening Version, which screens defendants to decide whether to administer the more in-depth LSI-R assessment (Lowenkamp, Lovins, & Latessa, 2009). Using a look-ahead period of 1.5 years, the study measured re-arrest rate and re-conviction rate, and found that a higher LSI-R score is positively correlated with future incarceration.

---

[9] In Philadelphia, for example, recidivism was being measured as re-arrest rate, and because of public opposition the sentencing commission began measuring it as subsequent conviction rate.

Interestingly, algorithmic risk assessments tend to find disparate validity levels when the same algorithm is used on racially distinct populations. Fass et al. in 2008 published validation data on the Level of Service Inventory - Revised (LSI-R) algorithm, as well as COMPAS (Fass, Heilbrun, DeMatteo, & Fretz, 2008). Using a dataset of 975 offenders released into the community between 1999-2002 from New Jersey, the measurement period was 12 months. The purpose of the study was to see whether these algorithms, trained on mostly white populations, are invalid for a population like New Jersey, which has has "substantial minority" representation in incarceration. The study finds "inconsistent validity when tested on ethnic/racial populations" (Fass et al., 2008, 1095), meaning the predictive validity may suffer as the result of differences between the training cohort used to develop the algorithm and the actual demographic breakdown of a jurisdiction. Demichele et al. in "The Public Safety Assessment: A Re-Validation" use data from Kentucky provided by the Laurence and John Arnold Foundation, which developed the PSA. The study measured actual failure-to-appear, new criminal activity, and new violent criminal activity before a trial. They found that the PSA exhibited broad validity, but found a discrepancy based on race (DeMichele et al., 2018).

Beyond recidivism, a few studies have focused on the relationship between risk assessment-driven decisions and other life outcomes, including earnings and family life. Bruce Western and Sara McLanahan in 2000 published a study entitled "Fathers Behind Bars" that finds alarming impacts of incarceration on family life. A sentence to incarceration was found to lower the odds of parents living together by 50-70% (Western & McClanahan, 2000). Dobbie et al. published a study that demonstrated that pre-trial detention in Philadelphia on increased conviction rates, decreased future income projects and decreased the probability that defendants would receive government welfare benefits later in life (Dobbie, Goldin, & Yang, 2018). The Prison Policy Initiative reports an unemployment rate above 27% for formerly incarcerated people, and find a particularly pronounced effects of incarceration on employment prospects for women of color (Couloute & Kopf, 2018).

Given the deeply impactful nature of risk-based decisions, validation experiments are surprisingly limited in scope. The outcome variable - typically rearrests in a one or two-year window - fail to capture the many ways that a risk-assessment can impact an individual's family, employment, income, and attitudes - all of which may be relevant in considering recidivism. Perhaps more importantly, the various aspects of life impacted by detention are precisely the risk factors that may get picked up by a subsequent judicial decision.

By treating risk assessment as instantial and analyzing longitudinal effects of a single assignment of risk, validation experiments are only observing part of the picture. When we consider the tangible impacts of judicial decisions and relate these impacts to future decisions, we see that there are possible feedback effects in the criminal system. The dependence of subsequent judicial decisions on prior judicial decisions is rampant. Sentencing guidelines suggest (and often require) judges to give longer sentences to repeat offenders, for example. The very notion of responsivity in criminal treatment requires periodic assessments that determine the 'progress' or treatment effect over time for a given defender, and shape punishment accordingly. However, treatment of sequential risk-assessments and the possible harms of feedback is missing from a literature that has so exhaustively debated whether incarceration has a criminogenic effect.

This thesis will explore how compounding in criminal justice impacts defendants. The treatment of risk assessment as innocuous, objective, statistical prediction has clouded rigorous theoretical exploration of lifetime compounding in criminal punishment. Using data from Philadelphia, we find that higher confinement sentences significantly increase cumulative future incarceration sentences for defendants. Synthesizing data from Philadelphia with a theoretical understanding of feedback in algorithmic risk assessment, we will discuss implications for judges and defendants.

# Part II

# Decisions

Bail and Sentencing in Philadelphia

# Chapter 3

# Pretrial

The history of criminal punishment suggests that the principle of risk is more complex and elusive than practitioners imply. With evidence from defendants in Philadelphia, we can begin to ground risk and see empirically how people are impacted. Who is risky? Where does risk concentrate? How does risk spread? And why does risk always seem to persist, despite a massive prison complex?

## *Background*

The U.S. jails about half a million people who have not been convicted (Wagner & Sawyer, 2018). After an arrest, judges must decide whether defendants should be 'released on recognizance' (without payment, also known as ROR), conditionally released, or detained before their trial. Conditional releases define terms that individuals must abide by in order to be released from jail - these may range from drug treatment programs to supervision. Commonly, courts will choose to conditionally release defendants by requiring a bond (Criminal Justice Policy Program, 2016). Bonds can be unsecured, which means that the defendant will owe a certain monetary amount in the event that they fail to appear in court, or they can be secured, which means they must pay some amount (typically 10%) in bail, up-front. When a defendant is unable to post the cash needed for a secured bond, a bail bond agent may act as a surety on the bond, posting the full amount of money on the condition that the defendant pay a fee and sign over a number of rights and privileges.

Bond agents often also require that defendants sign over collateral to cover the full bail amount - this might take the form of a house, a car, or property of a consenting family member (Wykstra, 2018). Bond agents are given broad authority to arrest defendants, search their belongings, and surveil them for additional criminal activity. They can require require that clients check in regularly, keep a curfew, and hand over medical, social security and phone records. In many instances, bondsmen can jail defendants who fail to pay loan fees. Given that bail bond agents' authorities extend far beyond those of a typical consumer finance company, they have come under scrutiny for extortion (Silver-Greenberg & Dewan, 2018).

Bail has come under attack in recent years for a variety of reasons. The commercial bail bond industry is bringing in 2 billion dollars annually in profit (ACLU, 2019) and lobbying heavily to oppose calls for reform (Duncan, 2014). But the bulk of their exploitative behavior is affecting people who can afford bond fees - the poorest defendants who can't afford fees have even bleeker prospects. Numerous studies demonstrate significant causal evidence that pre-trial detention has serious and long-term harms on an individual. Sacks and Ackerman in 2012 find that detention destabilizes family, increases expected incarceration length, and increases the likeli-hood of conviction (Sacks & Ackerman, 2012). Dobbie et al. find similar results: With compromised bargaining power, defendants who are detained before their trial are more likely to enter plea deals and incur guilty dispositions (Dobbie et al., 2018). Gupta et al. find detention increases recidivism in Philadelphia (Gupta, Hansman, & Frenchman, 2016), and another study found similar results in Texas (Stevenson & Mayson, 2017, 672). In Philadelphia, over half of people detained pretrial would be able to leave prison for a deposit of $1,000 or less, and many of these defendants are 'low-risk' - 60% of those held over three days were charged with non-violent crimes, and 28% just had a misdemeanor charge (Stevenson, 2018, 2). Pretrial detention also increases expected court fees and sentence lengths (Stevenson, 2018). A recently published study by Arnold et al. in 2018 used data from Miami and Philadelphia to find that judges exhibit significant racial bias in pre-trial release decisions, measured using offense rates of marginal white and black defendants (Arnold et al., 2018).

Calls to reform the bail system have led a few states, including New Jersey, to completely do away with cash bail. In its place, many jurisdictions use the PSA, which is a widely-used risk-assessment algorithm for pre-trial judicial decisions. The potential for algorithms to predict non-compliance may allow more lenient pretrial release measures for those that have a high probability of good behavior. It may also remedy judicial biases. These two points are widely cited by proponents of algorithmic risk assessment. In this section, we will explore the ways that pre-trial decisions impact Philadelphia. We'll closely analyze the PSA algorithm, and see how it would direct treatment to Philadelphia residents. With a more concrete understanding of risk and its complex relationship to different defendants in different places and circumstances, we find that predictive risk assessment does not escape all of the problems that plague the system of bail.

# *Data*

Historical bail decisions are included in court case dockets that are created for defendants in municipal and criminal court. However, the bulk of data used in our study came from court docket summaries that did not include information about bail. Instead, bail decisions were scraped from the Philadelphia Court's processing website, where "New Criminal Filings" are listed for a week before being removed from the site. These filings come from preliminary arraignments in Philadelphia Municipal Court, which oversees all criminal cases before sending more serious cases to the court of Common Pleas. Defendant name, age, zip code, charge, filing date, representation type, and custody information were listed, where applicable. In addition, bail date, bail status (posted/set/denied), bail type (secured/unsecured), bail amount, and outstanding bail amount are available. The defendant's court docket summaries are publicly accessible, so we were able to find extensive information about criminal history.

Decisions between February 2, 2019 and April 3, 2019 were recorded. Of course, the recency of the data available does not permit analysis of long-term outcomes resulting from bail decisions. But the data do provide robust information about the geographical, social and urban factors that surround bail decisions and the principle of risk more broadly. In all, 5611 observations were recorded, and, removing null values and repeat entries, we work with n=4889.

To analyze the tangible, geographical, and demographic information they may be encoded in risk scoring, we retro-actively compute the Public Safety Assessment risk scores for each individual. The algorithm relies on nine simple features about a

defendant:

> Age at Current Arrest
> Current Violent Offense
> Pending Charge at Time of Offense
> Prior Misdemeanor Conviction
> Prior Felony Conviction
> Number of Prior Violent Convictions
> Number of FTAs in the past 2 years
> Number of FTAs older than 2 years
> Prior sentence to Incarceration[1]

Each risk factor was calculated from docket summaries. As new states often have to adapt their individual way of recording data to adhere to the PSA's guidelines, signficant manipulation was necessary to turn Pennsylvania's docket summary sheets into PSA risk scores.

Below are summary statistics for variables used for analysis. In this section, we discuss simple findings from bail decisions in Philadelphia. These preliminary findings will motivate the more involved quantitative analyses in the remaining chapters.

---

[1] See Laurence and John Arnold Foundation, Risk Factors and Formula on the PSA Website.

Table 3.1: Summary Statistics for Philadelphia Municipal Court Filings

|  | mean | std | min | 25% | 50% | 75% | max |
|---|---|---|---|---|---|---|---|
| **Demographics:** | | | | | | | |
| Male | 0.82513 | 0.380 | 0.0 | 1.0 | 1.0 | 1.0 | 1.0 |
| Black | 0.59354 | 0.491 | 0.0 | 0.0 | 1.0 | 1.0 | 1.0 |
| Age | 33.3194 | 11.719 | 11. | 24. | 31. | 40. | 78. |
| **Risk Factors:** | | | | | | | |
| FTA Score | 1.73034 | 0.761 | 1.0 | 1.0 | 2.0 | 2.0 | 6.0 |
| NCA Score | 2.60429 | 1.331 | 1.0 | 1.0 | 3.0 | 4.0 | 6.0 |
| NVCA Score | 0.19244 | 0.394 | 0.0 | 0.0 | 0.0 | 0.0 | 1.0 |
| Num Prior Arrests | 5.25393 | 7.343 | 0.0 | 1.0 | 3.0 | 8.0 | 155.0 |
| Prior Misdemeanor | 0.48417 | 0.500 | 0.0 | 0.0 | 0.0 | 1.0 | 1.0 |
| Prior Felony | 0.36466 | 0.481 | 0.0 | 0.0 | 0.0 | 1.0 | 1.0 |
| Num Prior Crimes | 2.56302 | 3.553 | 0.0 | 0.0 | 1.0 | 4.0 | 49.0 |
| Num Prior Vio Crimes | 0.79265 | 1.468 | 0.0 | 0.0 | 0.0 | 1.0 | 16.0 |
| Prior Incarceration | 0.44351 | 0.497 | 0.0 | 0.0 | 0.0 | 1.0 | 1.0 |
| **Current Charge Info:** | | | | | | | |
| Current Vio. Charge | 0.45026 | 0.498 | 0.0 | 0.0 | 0.0 | 1.0 | 1.0 |
| Public/no lawyer | 0.94239 | 0.233 | 0.0 | 1.0 | 1.0 | 1.0 | 1.0 |
| **Bail Info:** | | | | | | | |
| Released Pre-Trial | 0.71542 | 0.451 | 0.0 | 0.0 | 1.0 | 1.0 | 1.0 |
| ROR/unsecured/cond | 0.56711 | 0.496 | 0.0 | 0.0 | 1.0 | 1.0 | 1.0 |
| Bail Amount | 70033.6 | 1.01M | 0.0 | 0.0 | 2.5K | 10K | 30M |
| Outstanding Bail Amt | 593.473 | 11510 | 0.0 | 0.0 | 0.0 | 0.0 | 750K |

## *Who Encounters the Criminal System?*

Municipal Court data on defendants arrested in Philadelphia represent a cohort that is far from randomly sampled among the larger Philadelphia population. We know this to be the case - poverty and crime are connected issues, and advancements in quantitative social science have begun describing these social ills using using networked and spatial quantitative approaches (Graif, Gladfelter, & Matthews, 2014).

Our data suggest that 59.4% of people arrested and charged with crimes in Philadelphia are black, even though the city is just 44.1% black, as of 2015 (Otterbein, 2015). Men account for 91.7% of arrests with preliminary arraignments. 60% had a former crime of some sort, and 44% have been sentenced to incarceration before. 94.3% of people had either "Defender Association of America" or "None" listed for legal defense - meaning that the vast majority of people do not get help from a lawyer when navigating the pretrial process, answering police and judicial questioning, and making decisions about bail. This statistic implies that folks who get arrested tend to have fewer resources to afford a private attorney.

With some understanding of the population that gets arrested and charged by police and criminal courts, we turn our attention to outcomes - in this case, bail decisions. Of those who get arrested, 42.6% are released on recognizance, 42.5% are offered cash bail, 12% are allowed to leave with an unsecured bail bond, and 2.2% are given non-monetary conditional release. Finally, from bail status, we can see that 28.5% of arrested people are kept in detention for some time. With a very small fraction (0.7%) denied bail, the rest (34.9%) are unable or unwilling to pay the money.

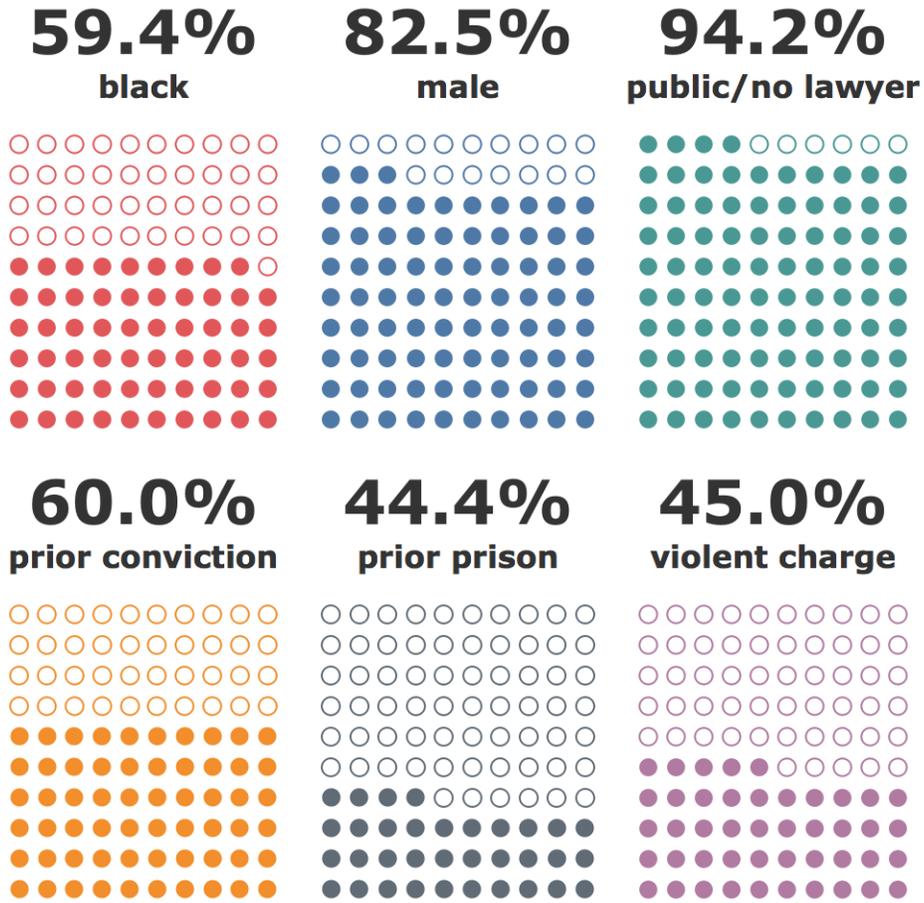

Figure 3.1: Population Statistics on Criminal Charges in Philadelphia

It appears from our Philadelphia dataset that a significant number of people are unable to pay bail, even at amounts that are quite low. While a smaller proportion of people are detained for the lowest bail amounts than the largest, a significant proportion of defendants cannot post bail at every bail amount. For cash bail amounts under $10,000, defendants would only have to post $1,000, and a bail bonds could be cheaper than $100. Most likely, defendants with bail this low were arrested for misdemeanors or non-violent charges, and are considered relatively low-risk. Yet, in just two months, we observe over a thousand people who are kept in prison because they are unable to afford bail. These people are more likely to enter plea bargains and accept guilty sentences, even if they are not guilty (Dobbie et al., 2018). Now imagine a risk assessment algorithm uses these individuals as data points - their higher susceptibility to a guilty disposition will train algorithms to identify people like them, fueling a cycle of risk-labelling that may be inaccurate, biased and costly to the state.

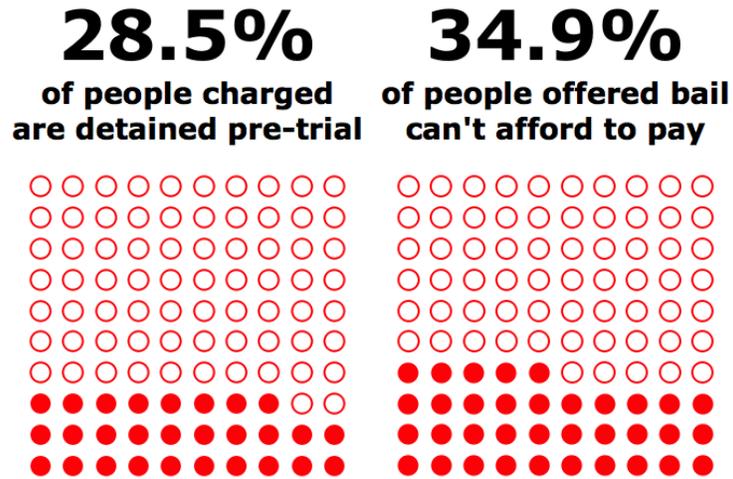

Figure 3.2: Bail Outcomes for People Charged in Philadelphia

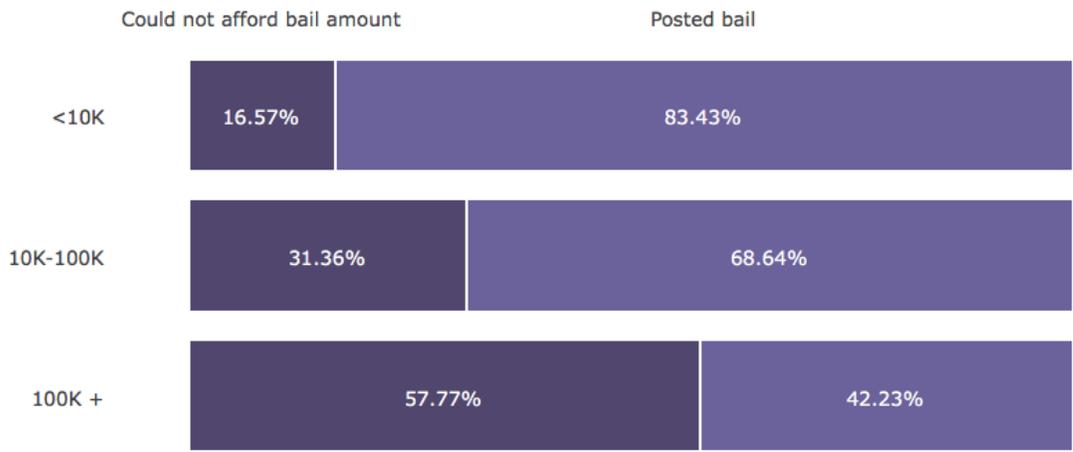

Figure 3.3: Ability to post bail at different ranges

Visualizing data on Philadelphia suggests that inequity and injustice is persistent. It is widely known that black Americans are over-represented in the U.S. prison system, compared to their population level in America. But a closer look shows that the cause for disparity is not a single decision - as defendants move through the pre-trial process, they are separated by factors like race and ability to pay money. Whereas 51.4% of black people arrested are released without having to pay cash up-front, the rate is 72.64% for everyone else. Of those offered cash bail, only 34.87% could afford to post bail, and the rest had to spend time in jail for at least a night. This rate also shows racial disparity - 33.96% of black defendants could pay bail, compared to 36.68% of other defendants. Ultimately, 31.77% of blacks are detained before their trial, compared to just 22.31% for others. These numbers indicate the broader point that judicial decision-making happens *sequentially*, and is closely connected to a host of other human decisions. Inequality does not stem from a single racist judge, or an algorithm that has not been trained correctly - it is the result of *systemic* forces that operate subtly over many decisions, events, and circumstances in society.

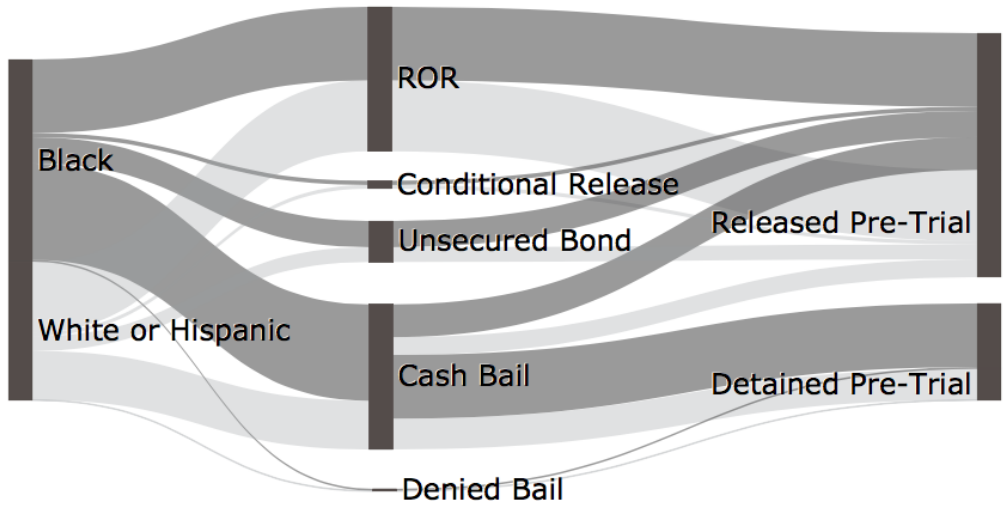

Figure 3.4: Bail decisions and outcomes by race

## *Mapping Risk*

With the current state of bail decisions in Philadelphia, reform is appealing to many across the political spectrum. In many ways, the glaring issues with bail have made risk assessments seem more palatable; made evident by the risk assessment developers who point to their ability to improve upon current judicial decisions. Are risk-driven, algorithmic policies exempt from the inequitable treatment that many have found in the cash bail system? Or does risk itself serve as a way to segregate along economic, racial and geographical lines?

With these questions in mind, we take a look at the risk level of defendants, scored using the Public Safety Assessment (PSA) algorithm. Though Philadelphia does not use the PSA to make pretrial decisions, we calculate the PSA on each defendant we observe, in order to see how risk is distributed across populations and locations. The PSA uses nine factors - all seemingly quite innocuous - to produce three scores. The scores correspond to a defendant's likelihood of failing to appear in court, committing a new criminal activity, and committing a new violent criminal activity.

The maps attached show various indicators by zip code in Philadelphia. The top row of maps contain demographic information about the observed population and the neighborhoods. The second row provide some insight about the realities of crime, arrests, and detention rates for people from each neighborhood. The third row of maps measure some of the decisions that are made by judges and law enforcement - in this case, we show arrests that led to charges, proportion of defendants released pre-trial, and cash bail levels for each zip code. Finally, the bottom three maps portray the three Public Safety Assessment risk scores.

We discuss a few general insights here. These insights are broad and qualitative in nature. They motivate further research into entrenched biases, accumulating disadvantage and racial disparities.

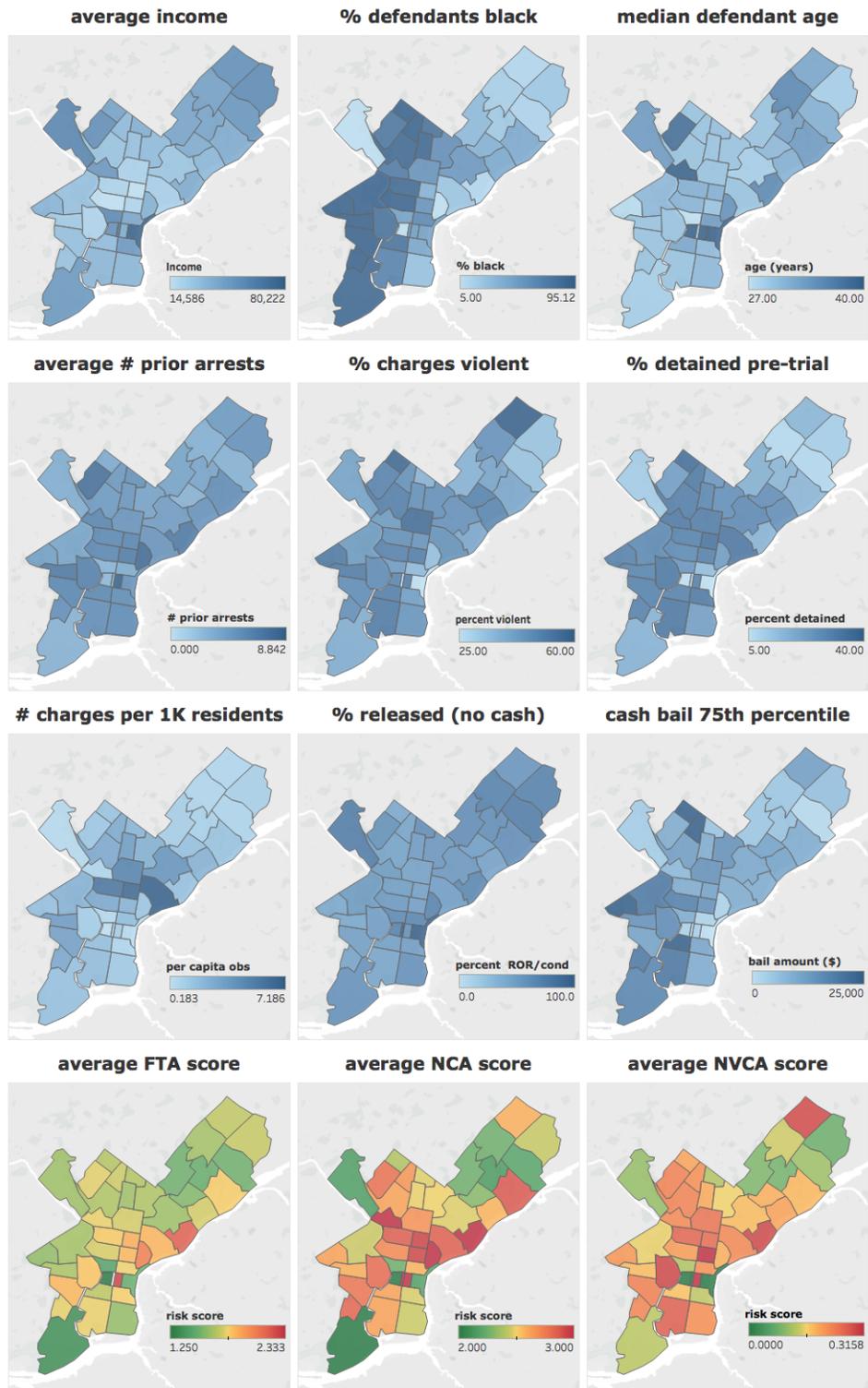

Figure 3.5: Mapping Risk in Philadelphia

It appears from the maps produced in Figure 4.1 that economic factors track geographically with criminal treatment, judicial decisions and risk level. High-income neighborhoods in the downtown area and on the outskirts of the city have low rates of arrest and charge. They also tend to be identified as low-risk. It also seems that higher-income neighborhoods are, on average, getting lower bail amounts, although this may be explained by lower violence rates in these neighborhoods.

Race also seems to overlap with certain other maps. Neighborhoods with the highest proportion of violent crime charges tend to have more black defendants. Also, it appears that race tracks with pretrial detention, which corroborates our earlier findings people of color are detained disproportionately compared to whites.

Finally, there appears to be a resemblance between the NCA risk map and historical arrest frequency, by zip code. Increased arrest rates in certain neighborhoods - often those that have lower income levels and higher black populations - may influence the factors that determine level of risk for future crime. In the coming chapters, we will discuss the possibility of historical dependence in more detail. But the relevance of geography is central to these broader questions in criminal policy and policing in cities.

## *Discussion*

Philadelphia's pre-trial process is far from perfect. Many defendants are being held in jail not because of their conduct but because of their ability to pay a small amount of money. Pre-trial detention profoundly impacts case outcomes, employment, and personal life. As such, calls for reform argue that ability-to-pay is not an appropriate basis on which to incarcerate some and release others.[2]

Black defendants from poorer neighborhoods are more likely to be held in jail before their trial. Return on Recognizance and conditional release appear to be much more common in more affluent neighborhoods, where violent crime is lower. Historical arrests and criminal histories imply that these patterns are entrenched and systemic, rather than the product of individual judicial decisions.

For those who concerned about the unequal treatment afforded because of America's bail system, replacing it with an algorithm that appears to treat everyone equitably seems like a good idea. However, a closer look at the pretrial population in Philadelphia suggests that assessing risk - whether by algorithm or by judicial discretion - may carry the same problematic inequities that characterize the bail system. The New Criminal Activity PSA score seems to be highest, on average, for defendants that come from neighborhoods with high historical arrest rates. These neighborhoods

---

[2] The ACLU's page on bail reform states that American's current system of bail is unconstitutional, as it violates due process and does not afford equal protection under the fourteenth amendment. See (ACLU, 2019).

tend to have the lower income levels, and commonly have more black residents.

These findings suggest that risk is not a designation that is inherent or fundamental. Instead, risk assessment tools draw information from variables embedded in complex web of choices, labels and historical realities. While replacing bail may significantly improve equal treatment, basing new decision-making processes on risk may continue to entrench some of the inequity that the reform aims to remedy.

# Chapter 4

# Sentencing

## *Background*

Pennsylvania is one of just a few states with a sentencing procedure that requires algorithmic risk assessment (Monahan & Skeem, 2016).This requirement began this year after a decade of contentious development. Tasked with developing the risk tool, PA's Sentencing Commission has reported that the algorithm is not the sole determinant of sentence length, and instead will be used only to determine those particularly high- and low-risk defenders, who judges may want to review with increased discretion. However, the algorithm's mandate, design and proposed use all fit squarely within the logic of selective incapacitation. In a report detailing the algorithm's risk factor inputs and formulas, the Pennsylvania Sentencing Commission describes the algorithm:

> The recommendation for additional information should apply to offenders determined to be at high risk of general recidivism or at low risk of general recidivism . . . This targeting of cases for additional information is consistent with the core principles of offender risk management: match the level of service to the offender's risk to recidivate; assess criminogenic needs and target them in treatment; and structure the sentence to address the learning style, motivation, abilities, and strengths of the offender.(Pennsylvania Sentencing Commission, 2018, 5)

"Level of Service" is a commonly used term that refers to the severity of sentence that is given to an offender. The quote above is undoubtedly punitive in its portrayal of criminal treatment - assessing risk to society comes first, and is still a central tenet to the intervention strategy. The quote does not reference any individual's past choices, but is instead derived directly from a defender's 'risk to recidivate. Additionally, highlighting that the tool is used most for only the highest and lowest-risk offenders, the Sentencing Commission follows a pattern of many other practitioners including Lowenkamp et al. who advocate separating convicted individuals to different facilities based on their level of risk (Lowenkamp, Latessa, & Holsinger, 2006).

Also missing from the risk assessment algorithm's mandate is any consideration of the *effect* of disparate sentencing decisions on an individual's future of criminal behavior and incarceration. By treating recidivism as an exogenous event, and purely aiming to predict risk, the commission ignores the potential that sentencing higher-risk people may cause more severe crime and fuel further incarceration. This effect, known as the criminogenic effect of incarceration, has been a discussion of social theorists for many years.

Different theories of criminal punishment frame the goals of incarceration differently - some emphasize it's deterrent effect on potential offenders. Others believe incarceration should be retributive, and should counterbalance social wrongs that have been committed. Proponents of rehabilitation theory believe that correctional facilities should address criminogenic needs of an offender. An understanding of what prison does to defendant's in the status quo is of course necessary to characterize incarceration and understand where it can be a useful tool in criminal punishment.

For this reason, theorists have for a long time debated whether incarceration empirically has a criminogenic, null, or deterrent effect on future crime. Camp et al. in 2005 find no criminogenic affect among 561 inmates in California with the 'same level of risk' who were distributed between Level I and Level III facilities - both were equally likely to be punished for misconduct in prison (Camp & Gaes, 2005). Bhati et al. in 2007 attempt to estimate the impact of incarceration on subsequent offending trajectories, and find little criminogenic effect - the bulk of subsequent incapacitation came from some sort of violation of the terms of incarceration, such as parole (Bhati & Piquero, 2007). Nagin et al. in 2009 also observe a null or mildly criminogenic effect on future criminal behavior (Nagin, Cullen, & Jonson, 2009). Vieratis et al., using panel data over 30 years in 46 states, find a population deterrent effect of increased incarceration rates, but also find that increased prison release rates lead to higher rates of crime incidents, on average (Vieraitis, Kovandzic, & Marvell, 2007). Harding et al. in 2017 analyze the effects of imprisonment on felony convicts in Michigan and, using randomized judges to establish causal inference, find that a prison sentence increases the probability of subsequent imprisonment by 18-19% (Harding, Morenoff, Nguyen, & Bushway, 2017).

The jury is out, so to speak, on the deterrent and criminalizing effects of prison. Thus, understanding whether a criminogenic effect exists in Philadelphia should be very relevant to an overhaul of sentencing guidelines. Instead, Pennsylvania joins other states in treating crime as external and inevitable; perhaps predictable but otherwise uncontrollable.

We analyze the question of criminogenic effects in sentencing by looking at Court of Common Pleas sentences in 2011. We wish to understand whether differences in sentencing that is not attributable to criminal risk or behavior has a long-term impact on cumulative prison time incurred via subsequent sentencing.

## *Data*

Philadelphia has made court summary documents filed since 2007 publicly available. These documents are created or updated each time a defendant has a preliminary arraignment subsequent to arrest. Preliminary arraignments, which typically occur a few hours after an arrest, allow a defendant to be notified of charges being brought, the date of a preliminary hearing, and information about bail. The Philadelphia Municipal Court has jurisdiction over all preliminary arraignment hearings, misdemeanor court trials, and preliminary hearings for felony trials. Felony trials and other more serious trials are heard in the Court of Common Pleas.[1]

To analyze whether subsequent incarceration has a feedback-effect on future incarceration rates, we limit our sample to defendants who have been convicted in the Court of Common Pleas. Scraping the first 12,066 case summaries filed in 2011, we obtain demographic information, historical arrest and court outcomes, crime severity, sentencing, and future court information to the present. Up-to-date docket summaries contain information about encounters with any court system in Philadelphia, as well as migrated cases from other jurisdictions.

Bail information is not readily accessible from docket summaries. However, when a defendant fails to appear at their trial, a bench warrant is issued and the status change is recorded in docket summaries. Using the thorough historical information in docket summaries, we retroactively compute PSA scores for defendants, as we did in our analysis of bail decisions.

---

[1] The following URL provides portal access to any docket summary by docket number: *https://ujsportal.pacourts.us/DocketSheets/CP.aspx*.

Table 4.1: Summary Statistics for Philadelphia Court of Common Pleas Cases

|  | mean | std | min | 25% | 50% | 75% | max |
|---|---|---|---|---|---|---|---|
| **Treatment Variable:** | | | | | | | |
| confinement_max (days) | 261.13 | 298.98 | 0.0 | 0.0 | 182 | 365 | 1095 |
| **Risk Factors:** | | | | | | | |
| fta_score | 1.8711 | 0.7320 | 1.0 | 1.0 | 2.0 | 2.0 | 5.0 |
| nca_score | 2.9260 | 1.3587 | 1.0 | 2.0 | 3.0 | 4.0 | 6.0 |
| nvca_score | 0.2671 | 0.4425 | 0.0 | 0.0 | 0.0 | 1.0 | 1.0 |
| prior_m | 0.4243 | 0.4943 | 0.0 | 0.0 | 0.0 | 1.0 | 1.0 |
| prior_f | 0.3496 | 0.4769 | 0.0 | 0.0 | 0.0 | 1.0 | 1.0 |
| plea_flag | 0.8348 | 0.3714 | 0.0 | 1.0 | 1.0 | 1.0 | 1.0 |
| number_prior_crimes | 2.4299 | 3.0812 | 0.0 | 0.0 | 1.0 | 4.0 | 29.0 |
| num_prior_violent_crimes | 0.8573 | 1.5220 | 0.0 | 0.0 | 0.0 | 1.0 | 14.0 |
| prior_incarceration_flag | 0.5084 | 0.5000 | 0.0 | 0.0 | 1.0 | 1.0 | 1.0 |
| num_prior_arrests | 5.5725 | 6.6846 | 0.0 | 1.0 | 3.0 | 8.0 | 60.0 |
| **Demographics:** | | | | | | | |
| age (days) | 12869. | 4026.9 | 4411 | 9739 | 1.2K | 15524 | 28360 |
| male_flag | 0.8887 | 0.3146 | 0.0 | 1.0 | 1.0 | 1.0 | 1.0 |
| black_flag | 0.6887 | 0.4631 | 0.0 | 0.0 | 1.0 | 1.0 | 1.0 |
| **Disposition Severity:** | | | | | | | |
| felony_flag | 0.7826 | 0.4125 | 0.0 | 1.0 | 1.0 | 1.0 | 1.0 |
| misdemeanor_flag | 0.3503 | 0.4771 | 0.0 | 0.0 | 0.0 | 1.0 | 1.0 |
| degree | 1.1762 | 1.1813 | 0.0 | 0.0 | 1.0 | 2.0 | 3.0 |
| (felony_flag)(degree) | 0.9186 | 1.2086 | 0.0 | 0.0 | 0.0 | 2.0 | 3.0 |
| (misdemeanor_flag)(degree) | 0.4684 | 0.8630 | 0.0 | 0.0 | 0.0 | 1.0 | 3.0 |
| count_guilty_charges | 1.7067 | 1.0492 | 1.0 | 1.0 | 1.0 | 2.0 | 12.0 |
| current_violent_charge | 0.5982 | 0.4903 | 0.0 | 0.0 | 1.0 | 1.0 | 1.0 |

## *Testing Criminogenic Effects of Incarceration*

Using data from Philadelphia, we can explore the impact of incarceration on defendants' futures. Specifically, we aim to explore the question of dependence in sequential decisions in criminal justice. Given a defendant with a certain propensity to violate the law, how might detention for a certain amount of time affect their expected future time in prison?

To test the impact of incarceration on life-courses in Philadelphia, we propose leveraging the fact that Philadelphia has not historically used algorithms to dictate sentencing decisions. Using risk factors *as controls* to compare between defendant outcomes, we analyze see how disparate sentence lengths may impact future incarceration. In other words, we set out to answer the following question empirically: Given two defendants with identical risk factors, how are differences in prison sentencing associated with cumulative future incarceration rates, measured up to 2 years after release?

Our question departs from existing criminogenic research because of its focus on confinement lengths, rather than criminal behavior or probability of arrest. We take no stance on the moral quality of the behavior of defendants, or the accuracy of trials or arrests as proxies for delinquency. Instead, we simply wish to see how incapacitation as a social phenomenon may breed self-reinforcement and feedback.

Reverse-engineering the PSA's three predictive scores - new criminal activity, new violent criminal activity, and failure to appear - as well as many underlying risk factors that are shared by algorithms across the country, we control for present risk of defendant at the time of sentencing. To control for the severity of a given crime, we include covariates to representing the typology of offense committed - felony and misdemeanor dummies, the 'degree' of the felony/misdemeanor, cross-terms, and the total number of guilty charges incurred.

We use a linear regression model with covariates X reported above. We aim to find the average incremental treatment effect of an extra day of sentenced prison time on the expected cumulative duration of prison sentences accrued until 2 years after the minimum sentenced time in prison. The treatment variable $x_1$ is measured using maximum sentences. We're fitting the following equation:

$$\hat{Y} = \beta^T X$$

$$Y := [2\_year\_cumulative\_incarceration\_mins]$$

$$\beta^T := [\beta_1, \beta_2, \beta_3, ...]$$

In the above equation, X refers to our observation matrix; each column represents our observation of a single defendant, and each row is a covariate listed in Table 4.1.

The potential for unobserved variable bias is important to note here, because judges may be seeing factors that are not reported in court docket summaries but may be relevant for sentencing. In particular, it is likely that judges cater sentences to different crimes that have the same grade, and may also cater sentences to particular combinations of multiple crimes that hold relevance for future incarceration prospects. To make sure our results are not representing our own shortcomings in modelling crime severity, we perform a second regression where we limit the sample to only defendants who commit the same crime, and who only are found guilty of that particular crime. We choose the most common crime in Philadelphia, "Manufacture,

Delivery, or Possession with Intent to Manufacture or Deliver" - a non-violent felony with degree = 0. For our second regression, we take out factors that have to do with current criminal severity, since everybody is convicted with the same crime.

## Results

A regression was performed for all cases in the Court of Common Pleas, and an additional regression was performed on only those cases which have an identical, single guilty disposition for drug dealing. With models described above, we test for the average incremental treatment effect of a day in prison on the expected cumulative length of prison sentences, measured until two years after the minimum prison sentence. Results are reported below.

Table 1: Criminogenic Effect of Confinement in Philadelphia

|  | *Dependent Variable:* | |
|---|---|---|
|  | 2-year minimum cumulative sentence | |
|  | All charges | Only Manuf./Possess./Deliv. |
| **Treatment Variable:** | | |
| confinement_max | 0.1286*** | 0.0938** |
|  | (0.0249) | (0.0451) |
| **Risk Factors:** | | |
| fta_score | -17.4921 | -103.3698*** |
|  | (21.7752) | (38.4002) |
| nca_score | 17.8048 | 66.3430*** |
|  | (13.5443) | (24.6393) |
| nvca_score | -12.3632 | 47.1921 |
|  | (23.4130) | (39.2216) |
| number_prior_crimes | 0.7477 | 0.1927 |
|  | (5.8815) | (11.2610) |
| number_prior_violent | -7.3534 | -19.4765 |
|  | (6.8279) | (13.1349) |
| prior_incarceration_flag | 3.5053 | -21.0556 |
|  | (23.1688) | (40.9863) |
| num_prior_arrests | 4.9809** | 4.6289 |
|  | (2.4096) | (5.1231) |
| prior_m | 9.4684 | 33.6151 |
|  | (18.8997) | (31.9502) |
| prior_f | 54.2776*** | 4.5463 |
|  | (18.3891) | (32.4370) |
| **Demographics:** | | |
| age | -0.0128*** | -0.0163*** |
|  | (0.0021) | (0.0041) |
| male_flag | 45.5795** | 65.2217 |
|  | (21.9017) | (52.8662) |
| black_flag | -7.8428 | -12.8142 |
|  | (14.6471) | (25.5853) |
| plea_flag | -70.8348*** | -155.8573*** |
|  | (19.8014) | (56.1067) |
| **Current Crime Severity:** | | |
| felony_flag | -64.1872 | |
|  | (40.2028) | |
| misdemeanor_flag | -17.5152 | |
|  | (35.4556) | |
| degree | -33.2990 | |
|  | (24.0823) | |
| (felony_flag)(degree) | 40.5926* | |
|  | (23.1846) | |
| (misdemeanor_flag)(degree) | 19.8258 | |
|  | (18.0401) | |
| count_guilty_charges | -19.5394** | |
|  | (7.6839) | |
| current_violent_charge | 29.0483 | |
|  | (17.7124) | |
| $N$ | 6215 | 1473 |
| $R^2$ | 0.008 | 0.033 |
| Adjusted $R^2$ | 0.007 | 0.023 |
| $F$-statistic | 7.323*** | 3.521*** |

*Standard errors in parentheses.*     *$p < .1$, **$p < .05$, ***$p < .01$

Regression results indicate that an additional day of sentencing is associated with 0.129 more days in prison sentences accrued two years after release, on average. For non-violent drug felony offenders, the estimated effect of incarceration is 0.094 extra days of prison time, on average. The regression that included all types of crime was statistically significant with $p < 0.01$, whereas the drug-only regression was statistically significant at $p < 0.05$.

While the results do provide evidence of a criminogenic impact of incarceration, it's important to note the possible alternative explanations for the observed treatment effect. First, unobserved variables might be influencing judge decisions. If judges use factors that were not controlled for and statistically correlate with future crime rates, we might observe the correlation in sentencing, which would suggest a causal relationship that is not only explained by differences in sentencing rates. We included the second regression because we were concerned that this bias might exhibit itself in the broad sentencing regression. The fact that there appears to be strong evidence of a criminogenic effect in both regressions is promising, but there may be other unobserved variables. One that may currently influence judicial decisions, since it is being adopted as part of Philadelphia's new sentencing tools, is juvenile delinquency history. Unless juveniles were tried in adult court, their record in inaccessible. While such a practice on face value seems to confirm our claim that sequential decisions in criminal justice compound (and are highly sensitive to inital conditions), being able to include juvenile information as another risk control would improve our confidence in concluding a criminogenic effect of incapacitation.

## *Some Evidence of Divergence*

The regression results reported above use a two-year measurement window for outcomes. To explore the more dynamic effects of incarceration, we calculate the population-wide cumulative sentences for 1, 2, 3, 4, and 5 year windows.

For both regressions conducted - all crimes and only single drug dealing convictions - we found that the estimated effect of incarceration on subsequent incarceration lowered year-by-year, and also decreased in statistical significance. This is to be expected, because people are more likely to reoffend in the transitionary period subsequent to release. Below, we plot sentence totals over time for white and black defendants. The plots show a persistent gap in conviction rates, where formerly-incarcerated black people are more likely to collect additional prison sentences than formerly-incarcerated whites. However, we can't be resoundingly confident of this difference from the data alone. An estimator for the difference year-by-year, as well as the standard error, is included below. Importantly, the error bars in the second plot represent a single standard error in either direction of our estimator, so the difference does not pass statistical confidence at the current number of observations.

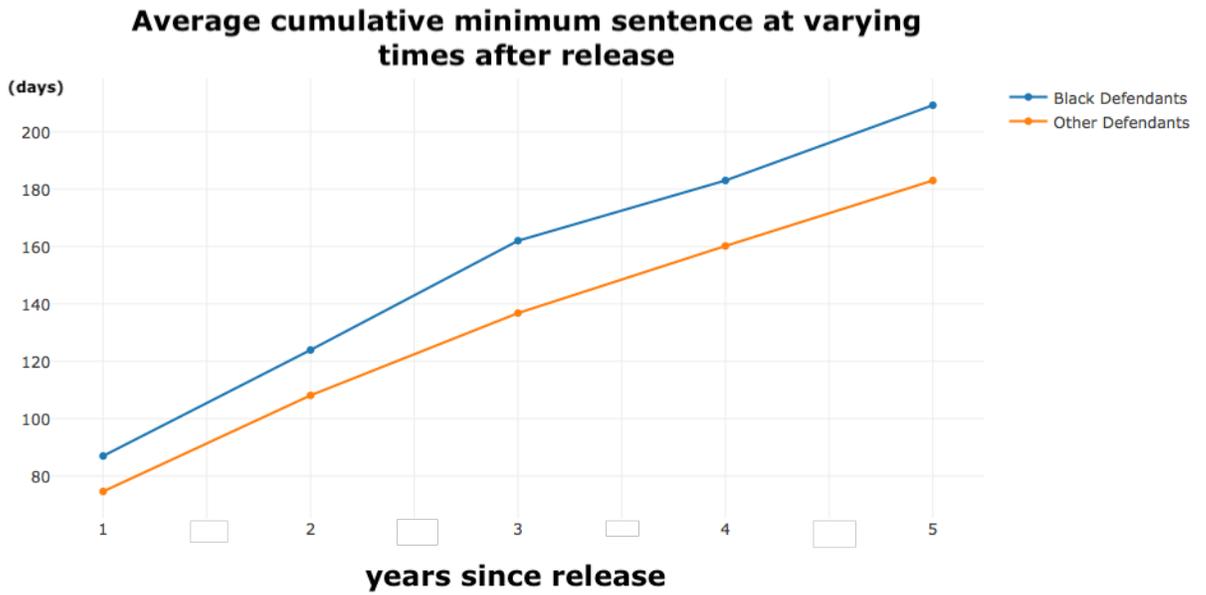

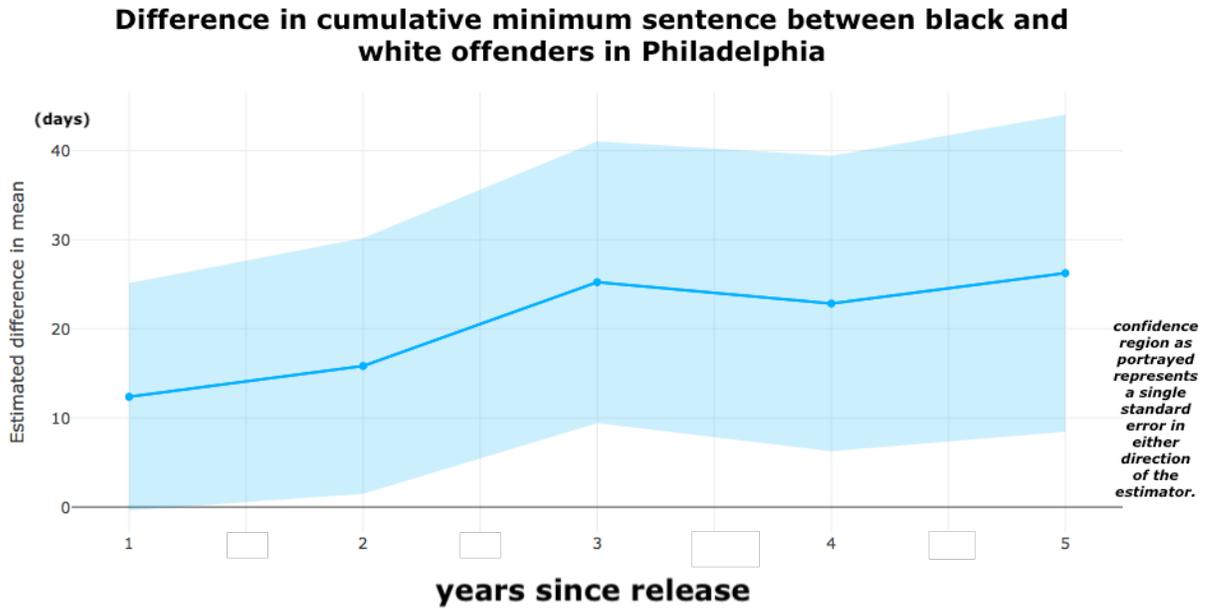

Figure 4.1: Potential Racial Divergence in Outcomes

# *Discussion*

Empirical results suggest that when a defendant is sentenced to an extra day in prison in Philadelphia, they can expect to spend more than one extra day in prison over the course of their lifetime. There are numerous explanations for why this may be the case, and there are numerous implications for policy-makers.

A criminogenic effect may result from peer networks that develop in prison. Some have found that exposure to people who are 'high risk' can increase crime (Lowenkamp et al., 2006). Other explanations include harmful effects of being labeled as a criminal for employment and family (Western & McClanahan, 2000). Higher sentences may also come with higher surveillance after release through programs like probation, which would lead to higher hit rates for former offenders, even if they do not commit crimes at higher rates. Regardless, significant statistical evidence suggests that this effect does exist broadly in Philadelphia.

The effect of prison time on future encounters with criminal punishment implies that algorithmic risk-assessment tools cannot be assessed using instantial experiments at one time in a defendant's life. Of course, it's important to mention that the study above *is* an instantial experiment at one point in a defendant's life. It also uses a risk-assessment tool to control for a defendant's propensity to be re-convicted. However, it is using these methods to explore how decisions impact defendants, not to understand a defendant's fundamental propensity to violate the law. If a single criminal sentence can impact a defendant's life outcome, subsequent sentences may *add* to this effect, and prison time can compound. Thus, we do not argue that the treatment effect encompasses the entire effect of criminal sentencing. Instead, it suggests that sentencing decisions may follow a compounding process, and disadvantage may accumulate for defendants over numerous encounters with the criminal system.

Our dataset suggests that defendants tried in Philadelphia's Court of Common Pleas can expect to be arrested more than two more times in the future, regardless of the number of times they've been arrested in the past. If larger sentences are associated with greater prison time, it is likely that longer sentences hold bearing on future risk assessment. A more severe sentence may lead parole officers to have more discretion over parolees. It may increase a defendant's association with other criminals. This kind of dependence between decisions is clear from sentencing tables and three-strikes rules, which recommend that judges give exaggerated sentences to repeat-offenders.

Since judicial decisions appear to feed into one another sequentially over a defendant's life time, it is important to consider models that encompass compounding effects. Risk assessment algorithms and validation experiments fail to adequately address the potential of feedback effects over time. Rigorously considering the impacts if dependent, sequential decisions will be necessary for any high-stakes algorithm that makes decisions temporally. In the forthcoming section, we explore the possibility of compounding disadvantage and model problematic effects that may arise, undetected by instantial validation techniques.

# Part III

# Impacts

The Dangers of Compounding Injustice

# Chapter 5

# Theoretical Modeling

Disadvantage can accumulate over time. The notion of compounding effects in decision-making is intuitive – discrimination is instantiated when somebody consciously discriminates, but the effects of discrimination are often felt when the bias is more insidious and systemic. For example, even if gender-based discrimination is nearly undetectable at a single stage in a company's hiring or promotion process, executive teams tend to show remarkably little diversity (Probert, 2005). Similar effects have been observed in education and wage rates, where a lifetime (or even inter-generational) time frame is needed to understand how bias becomes entrenched and can perpetuate over time.

Thus, statistical methods that try to find instances of discrimination may not capture biases that compound over repeated decisions. Another challenge for research is the difficulty of developing rigorous models of systemic effects. These processes can be highly complex because they involve information about history – something that traditional regression techniques lack. In a text entitled "Measuring Racial Discrimination" by the National Research Council in 2004, a chapter devoted to compounding effects concedes that the field is under-analyzed. The text observes, "Measures of discrimination that focus on episodic discrimination at a particular place and point in time may provide very limited information on the effect of dynamic, cumulative discrimination" (Paer, 2005, 226). As a result, more research is needed, despite modeling difficulties. The authors write:

> Very little research has attempted to model or estimate cumulative effects. In part, this is because modeling and estimating dynamic processes that occur over time can be extremely difficult. The difficulty is particularly great if one is trying to estimate causal effects over time. (Paer, 2005, 224)

Indeed, theorists have found that survey and panel experimentation usually have not been able to capture the accumulating disadvantage that can cyclically affect a group of people, or cause divergent levels of wealth or status in society (Lyons & Pettit, 2011). Instantiated experiments are unable to capture the dynamic nature of cumulative effects, and therefore often underestimate coefficients that determine measure of inequity.

What are dynamic effects, and how might they be occurring in criminal justice? If risk, as currently defined, compounds over time, is it a proper goal to cater punishment severity to risk? We will explore the theoretical underpinnings of risk. A rigorous treatment of dependence in sequential decision-making indicates that, indeed, compounding effects are possible and have the potential to lead to unexpected and unfair practices for certain defendants.

# *Context*

Predictive decisions commonly use variables that change over time. Risk assessment literature makes a designation between 'static' and 'dynamic' features(Picard-Fritsche et al., 2017), though even so-called static features include information about criminal history, which of course can changes over time. As algorithms are adopted at more stages throughout a defendant's life (arrest, bail, trial, sentencing, confinement, probation, parole, re-arrest), changes in defendant characteristics become entangled with criminal decision outcomes. Pretrial detention has been shown to increase the probability of conviction by lowering bargaining power, decrease employment opportunity, and decrease future government assistance (Dobbie et al., 2018). It also increases rates of re-arrest after disposition (Gupta et al., 2016), (Leslie & Pope, 2017). Downstream effects ranging from family stress to court fees have profound impacts on people's lives, and indeed, on features that are deemed 'risk factors' for defendants. A report from the Center for Court Innovation categorizes the most common factors used by algorithms today:

| | |
|---:|:---|
| Criminal History | Substance Abuse |
| School or Work Deficits | Antisocial Personality Pattern |
| Demographics | Leisure Activities |
| Family Dysfunction | Criminal Peer Networks |
| Antisocial Attitudes | Residential Instability[1] |

---

[1] See (Picard-Fritsche et al., 2017, 5-6).

Perhaps with the exception of demographics, each one of these factors is profoundly impacted by months of detention. Thus, in the event that numerous assessments are used consecutively on the same person, there may be unintended effects.

## *Modelling Sequential Risk Assessments*

We begin with a simple model of risk-needs driven decisions. Given that existing risk assessment services emphasize their wide applicability, some algorithms are adopted at numerous stages in criminal proceedings.[2] Other jurisdictions may use different assessments for policing, bail, sentencing and parole. Starting simple, we model risk assessments as instantaneous binary decisions that are separated in time. Each decision occurs sequentially, and the outcome is either "high risk" or "low risk", as visualized below.

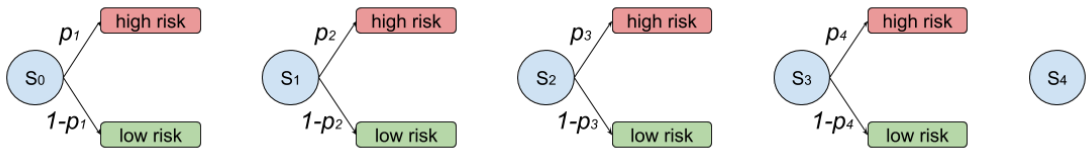

We assume here that risk assessments are conducted $n$ times throughout a person's

---

[2] An LSI brochure claims their algorithm is "proven to accurately predict recidivism, violence, and a large number of other relevant outcomes." It continues, "One of the most widely used instruments for the assessment of recidivism, the LS instruments are currently being used by probation, parole, community corrections, prisons, psychologists and mental health professionals." URL: https://issuu.com/mhs-assessments/docs/ls-cmi.lsi-r.brochure_insequence?e=20431871/45044118. See also COMPAS Case Supervision Review tool for repeatedly re-assessing risk during detention/proceedings. URL: http://www.northpointeinc.com/files/downloads/FAQ_Document.pdf.

life, and each decision $i \in \{1, 2, ..., N\}$ is a random variable denoted $X_i$, with:

$$X_i \in \begin{cases} 1, & \text{if defendant is classified high-risk} \\ 0, & \text{if defendant is classified low-risk} \end{cases}$$

We model each assessment using the current state of the world before decision $i$, denoted $S_{i-1}$, and the probability that a defendant will be designated high-risk, denoted $p_i$:

$$P(X_i = 1 | S_{i-1}) = p_i$$

The assessment is a random variable and not deterministic because risk assessment algorithms do not solely determine defendant outcomes - the ultimate decision is still up to a judge, who references the risk assessment score as part of the broader pre-trial policy decision.

We wish to explore the possibility that outcomes of assessments may impact and alter future assessments. As such, our model must enable us to analyze cases where the outcome variable $X_i$ may impact the probability of high-risk classification for $X_{i+1}, X_{i+2}, ..., X_N$. The probability of a high-risk classification at decision $i$ can thus be thought of as a function of some defendant information $D_i$ (gender, race, age) and the history prior decisions, $H_i$. We write the current state of beliefs at $i$ as $S_i = \{D_i, H_i\}$. We more accurately portray this dependence on the history of decisions as a branching process, rather than a sequence of decisions:

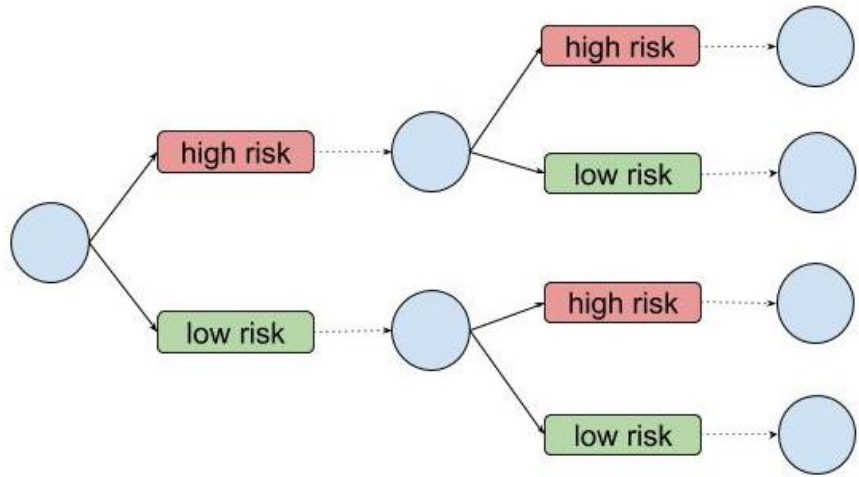

Every major risk assessment algorithm uses information about criminal history to assess risk. PSA, for example, measures a defendant's number of prior misdemeanors, felonies, convictions, and violent convictions.[3] These numbers add various point values to a risk assessment score, and a threshold value may determine pre-trial detention or cash bail amounts. Therefore, the PSA and most (if not all) other algorithms have a reinforcement effect. After an individual is convicted with a felony charge, every subsequent risk assessment for the rest of his life will use his criminal history to increase his risk score. Thus, initial assessments of risk can hold more 'weight' in determining lifetime treatment than later assessments. If a person is identified as high-risk in their first encounter with the criminal system, known effects on future crime rates, employment, family life, taxes, and other features will increase the likelihood of subsequent encounters.

---

[3] Public Safety Assessment: Risk Factors and Formula. URL: https://www.psapretrial.org/about/factors.

This property of *reinforcement* is key to modeling our system. The process is not Markovian: history matters, and our state of beliefs changes over time. Instead, we understand the changing effects of sequential risk-assessments as an Urn process, derived from the classic Pólya Urn model in mathematics (Pemantle et al., 2007).

## *Dependence and Reinforcement*

Let's say each risk assessment decision affects subsequent decisions as follows: If $X_{i-1}$ is the risk-assessment outcome for decision $i-1$, the subsequent probability of a high-risk decision $p_i$ is a weighted average between $p_{i-1}$, the prior probability, and $X_{i-1}$, the most recent classification:

$$p_i = p_{i-1}\left[\gamma_i\right] + X_{i-1}\left[1 - \gamma_i\right], \quad i \in \{2, ..., N\}, \quad \gamma_i \in [0, 1]$$

This means that we model updates in risk score by averaging the prior assumed risk and the outcome of a new assessment. The $X_{i-1}$ term can be thought of as the marginal effect of a new classification on defendant risk. To model reinforcement, we allow $\gamma_i$ to increase as $i$ increases, letting prior risk score $p_{i-1}$ hold more importance as a defendant is older and has more history. This should make intuitive sense - if a defendant has lived out most of his life with a certain propensity for criminal activity ('risk'), the effect of a new assessment should carry less weight.

Using the above intuition, we'll start by assuming the following relationship be-tween $\gamma_i$ and $i$ (the number of encounters with the criminal justice system):

$$\gamma_i = \frac{i}{i+1}$$

To understand the equation above, let's consider the value of $\gamma_i$ for varying $i$. In a first encounter with criminal courts where $i = 1$, we'd have $\gamma_1 = \frac{1}{2}$. Risk assessment outcome $X_1$ would thus have a very strong impact on future risk assessments. When $i$ is high, however, $\gamma_i$ approaches 1 and new assessments would diminish in weight. This is the reinforcement property we're seeking - the more decisions that go by, the less weighty they are in determining a person's lifetime experience with the state's criminal system.

Thus, our formula for $P(X_i|D, H_i)$ is:

$$P(X_i|p_{i-1}, X_{i-1}) = p_{i-1}\left[\frac{i}{i+1}\right] + X_{i-1}\left[\frac{1}{i+1}\right], \quad i \in \{2, ..., N\} \quad (5.1)$$

Let's assume temporarily that every defendant starts off with a probability of high-risk classification $p_1 = \frac{1}{2}$. We model the effect of sequential risk-assessments for different defendants by implementing our iterative equation. Below are sample paths for 5 defendants who are subject to ten periodic, evenly spaced assessments over time:

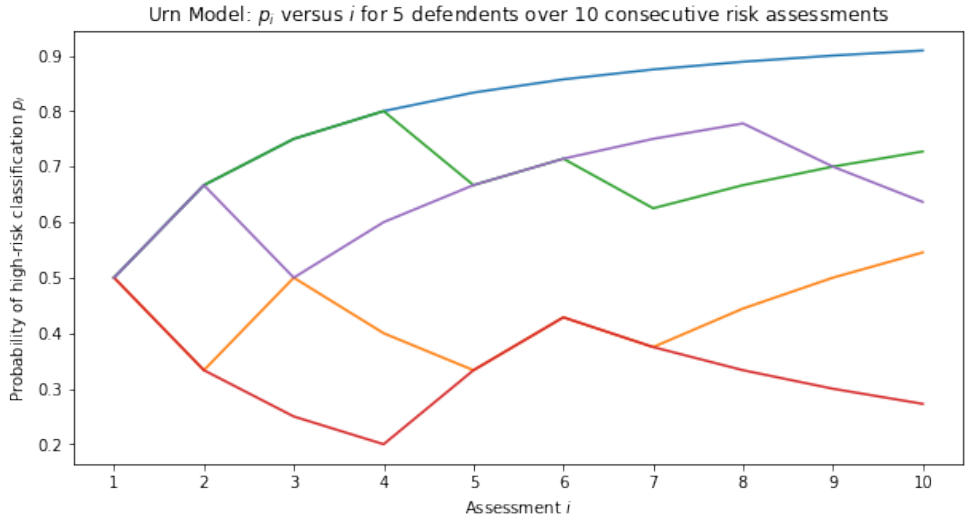

In the plot above, each color represents an individual who encounters criminal risk assessments throughout their life. Notice that this plot behaves in accordance with the reinforcement effect - initial assessments have large effects on $p_i$, and later assessments only marginally change the course of the risk level. Indeed, the for very large $i$ the risk level approaches a straight-line, meaning that the system reaches a stable propensity for criminal activity. Below are the paths of the same five defendants, this time over a total of 100 assessments (so 90 additional assessments):

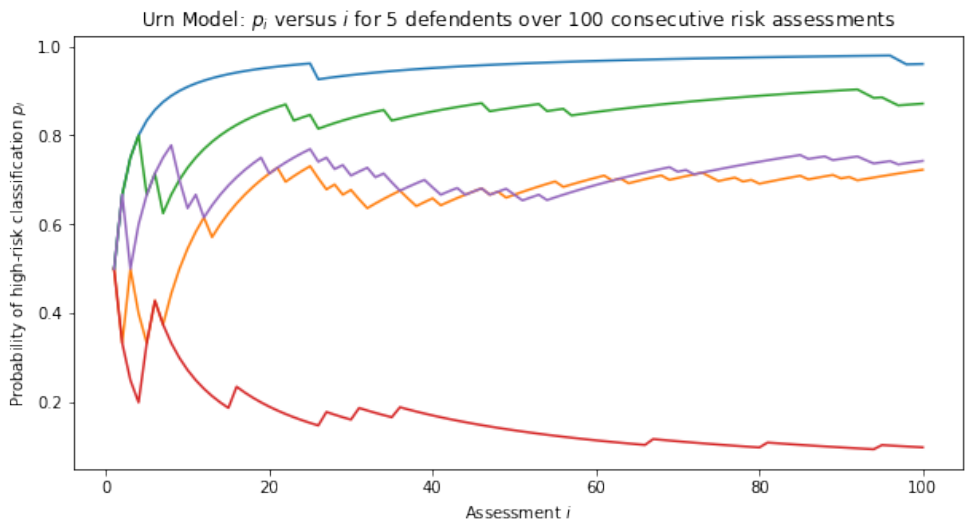

While it is unrealistic that a single person would have one hundred exactly evenly spaced and identical assessments throughout their lives, the behavior of our model seems to cohere with our knowledge of risk-assessments - their output impacts future assessments in a way that reinforces their classification. In other words, people detained after being identified as high-risk are more likely to re-offend, spend time in jail, have financial trouble, lose employment, or receive a guilty charge - all of which will affect their level of 'risk'.

# *Pòlya's Urn Generalization*

The model derived above is an Urn process. Borrowing a few theorems from probability theory, we can begin to understand the large-scale, long-term effects that might come about when algorithms are used consecutively throughout a person's life.

## Pólya's Urn Model

Pòlya's Urn is a classic model in probability theory, introduced by George Pòlya in an attempt to model infectious disease.(Pemantle et al., 2007, 5) The model describes path-dependent branching processes that are 'exchangeable', meaning the order of prior events does not matter.[4] The model asks what the long-term distribution of blue balls will be in the following random process:

---

[4] This is an assumption that may not hold true for our case, because many algorithms care about how *recent* a historical event took place. PSA, for example, cares about prior failures to appear in court in the past two years. However, for the most part, algorithms consider the aggregate number of historical events - number of prior felonies, misdemeanors, convictions, etc. These indicators are all *exchangeable* in the sense that it doesn't matter when in the defendant's life they occurred.

- An urn contains $R_i$ red balls and $B_i$ blue balls. Start at $i = 0$, with an initial mix of $R_0$ and $B_0$ balls.

- for iteration $i \in \{1, ..., N\}$:
    - Pick a ball randomly from the urn.
    - For the ball picked, return it and $k$ additional balls of the same color to the urn.

## Urn Equivalence to Risk Assessment Model

We can model reinforcement in algorithmic decision-making as an urn process. Our basic defendant model replicates exactly the basic Pòlya process with $R_0 = 1$, $B_0 = 1$, and $k = 1$. We derive the equivalence in the two processes below.

Denote the color of the ball selected by pick $i \in \{1, 2, ..., N\}$ as:

$$\tilde{X}_i \in \begin{cases} 1, & \textit{if blue ball is picked} \\ 0, & \textit{if red ball is picked} \end{cases}$$

Assuming each ball is picked with equal probability, the probability of picking blue in is given by:

$$P(\tilde{X}_i = 1) = \frac{B_{i-1}}{B_{i-1} + R_{i-1}}$$

The total number of ball in the urn is $n_i = R_i + B_i$. The probability of picking blue given all prior picks is denoted as $\tilde{p}_i$. We can always find $\tilde{p}_i$ by dividing the number of blue balls in the urn by the total number of balls. We've shown that $p_i = \frac{B_{i-1}}{n_{i-1}}$. After the $i^{th}$ pick, what will be the probability of picking blue? We inevitably add $k$ balls into the urn, so $n_i = n_{i-1} + k$. In the event that our pick is red, we still have $B_{i-1}$ blue balls, so the probability of picking blue decreases to $\frac{B_{i-1}}{n_{i-1}+k}$. If we do pick blue, however, the probability increases to $\frac{B_{i-1}+k}{n_{i-1}+k}$. Thus, the probability of picking blue on the $(i+1)^{th}$ pick, given $B_0, n_0$ and $\tilde{X}_1$, is:

$$\tilde{p}_{i+1} = \frac{B_{i-1} + \tilde{X}_i k}{n_{i-1} + k}$$

With a bit of algebra, we can define this probability in terms of the probability for the prior pick:

$$\tilde{p}_{i+1} = \frac{B_{i-1}}{n_{i-1} + k} + \tilde{X}_i \frac{k}{n_{i-1} + k} = \left[\frac{B_{i-1}}{n_{i-1}}\right] \frac{n_{i-1}}{n_{i-1} + k} + \tilde{X}_i \frac{k}{n_{i-1} + k}$$

$$\therefore \tilde{p}_{i+1} = \tilde{p}_i \frac{n_{i-1}}{n_{i-1} + k} + \tilde{X}_i \frac{k}{n_{i-1} + k}$$

When $k = 1$ and $R_0 = B_0 = 1$, how does $n_i$ behave? It starts at $n_0 = 2$, and after each pick it increments by $k = 1$. Thus, $n_i = 2 + i$. Equivalently, $n_{i-1} = 1 + i$, and $n_{i-2} = i$. Using the relationship derived above, a shift in index yields the probability of picking blue $\tilde{p}_i$ for $i \in \{2, ..., N\}$:

$$\tilde{p}_i = \tilde{p}_{i-1} \frac{n_{i-2}}{n_{i-2} + k} + \tilde{X}_{i-1} \frac{k}{n_{i-2} + k} = \tilde{p}_{i-1} \left[\frac{i}{i+1}\right] + \tilde{X}_{i-1} \left[\frac{1}{i+1}\right] \quad (5.2)$$

Notice the equivalence to equation 5.1. We've shown the probability for picking blue at each iteration of the classic Pólya Urn process exactly equals the probability of a high-risk classification in our simple model of sequential risk assessments, where $\tilde{p}_i = p_i$ and $\tilde{X}_i = X_i$.

## *Long-Run Behavior*

When we say that a sequence of random decisions might exhibit *reinforcement*, we now know that this means something deeper mathematically. Random processes with reinforcement behave in certain ways that might be problematic in the context of criminal policy. We have a general sense that algorithmic decisions in criminal justice impact defendants profoundly, and likely impact future encounters with law enforcement. Leveraging insights from probability theory, we can begin to understand the danger of policies that have compounding effects.

To start, we analyze the long-term treatment of individuals that are subject to sequential risk-based decisions. In Robin Pemantle's "A Survey of Random Processes with Reinforcement" (2006), the following theorem is reported about Pòlya's Urn process:

> Theorem 2.1: The random variable $p_i = \frac{B_i}{B_i + R_i}$ converges almost surely for large $i$ to a limit $P$. The distribution of $P$ is: $P \sim \beta(a,b)$ where $a = \frac{B_0}{k}$ and $b = \frac{R_0}{k}$. In the case where $a = b = 1$, the limit variable $P$ is uniform on $[0,1]$. (Pemantle et al., 2007)

Theorem 2.1 lays out how we can expect our modeled risk assessments to behave over many iterations. If one person undergoes risk assessments numerous times throughout their life, they may end up in radically different places depending on the risk-assessment outcome. They may be able to steer clear of subsequent confinement and re-arrest, or they may be continuously surveiled and repeatedly penalized by the state.

For a preliminary understanding of how inter-dependence in repeated risk assess-ments can impact a population, we use our initial modeling assumption that $p_1 = 0.5$ (so $B_0 = R_0$ and $a = b$), and imagine varying the parameter that determines the bearing of prior assessments on updated assessments, $k$ (which defines $\gamma$). If we decrease $k$ to 0.1 so that $a = b = \frac{B_0}{k} = 10$, we have the following long-term distribution for defendant risk:

Figure 5.1: PDF of long term risk level when $k = 0.1$

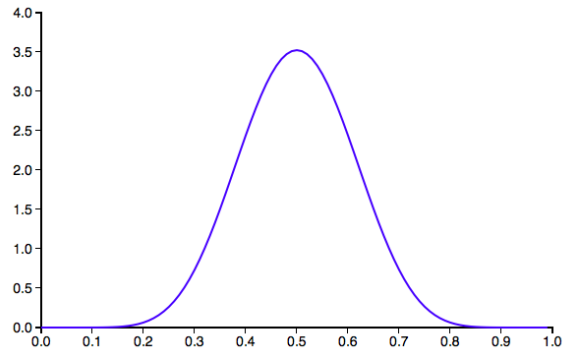

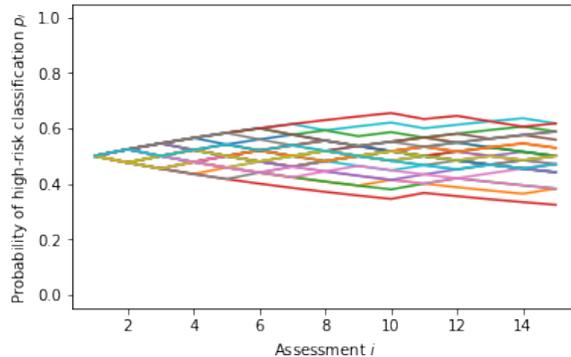

When decisions have little impact on people's lives (and potential subsequent risk assessments), we see consistency in long-term outcomes. Everyone starts with a risk score of 0.5, and all end up somewhere near there even after many assessments.

However, if algorithmic-driven decisions are more sensitive to the effect of prior decisions with $a = b = \frac{B_0}{k} = 0.1$, then we can see very problematic behavior in the long term:

Figure 5.2: PDF of long term risk level when $k = 10$

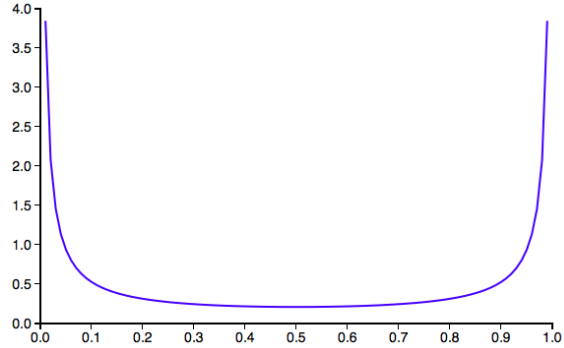

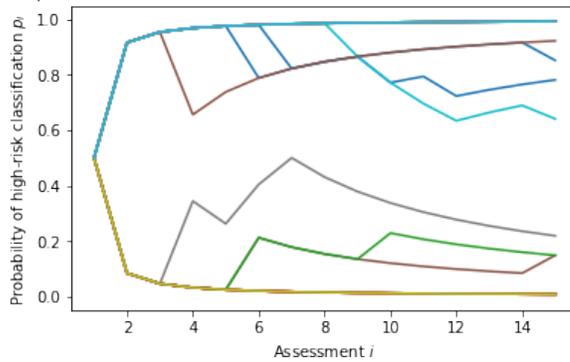

In this second case, we begin with defendants that are identical in attributes, with an initial probability of high-risk classification $p_1 = 0.5$. However, simply because of the effect of risk-based decision making, defendants end up with radically different risk levels, and are highly likely to be pushed to an extreme (no criminal risk, 0, and extreme criminal risk, 1).

Of course, these results are purely theoretical and do not come from real observed processes. But they motivate the importance of scrutinizing how algorithms are used in practice. Algorithms may be validated to ensure that biases are mitigated to a certain confidence threshold. But even tiny disparities in the system described by the second plot above can profoundly impact outcomes.

# *Modeling Inequality*

Many critics of risk assessment tools have expressed concern that these tools may encode biases that have historically characterized United States law enforcement. So far, our analysis of compounding effects has shown that these tools can lead to radically disparate treatment between people who began with the same risk factors. However, the analysis has not yet touched on existing and historical inequity. If a *biased* risk assessment tool were used, *and* it exhibited compounding effects, how might we expect bias to propagate over time? We can use our urn model to answer this question theoretically.[5]

## Disparate Initial Conditions

Risk assessment tools claim to add a level of consistency and 'objectivity' that judges lack without algorithmic assistance. Since judges have historically been biased in certain ways, many algorithmic tools boast that their improved accuracy can allow more people (of all groups) to leave detention pre-trial without increasing crime rates.

Even if we assume that our algorithm perfectly predicts risk and is able to eschew any kind of racially encoded bias, we know factually that risk is unevenly distributed across race.[6] A randomly selected black individual who finds himself arrested for a crime, therefore, is more likely to be labeled as high risk than an average white person in the same circumstances[7].

---

[5] (Kleinberg et al., 2017) discussees lowering the number of black people incarcerated as a potential goal for algorithmic criminal decisions.

[6] See (Harcourt, 2014).

[7] (Harcourt, 2008, The Virtues of Randomization) demonstrates that, as long as there is profiling, the arrested population will not accurately represent the true offending population demographically (absent perfect crime detection).

What are the long-term impacts of adopting algorithmic risk-assessments when risk is unevenly distributed across racial groups? How can our simple model of sequential risk assessments help us understand compounding effects and biased treatment?

Our first line of inquiry will look at the initial risk score that a defendant receives in a first encounter with the criminal justice system. Recalling our sequential decision-making model, we were able to describe the entire system with two quantities: the initial 'risk level' $p_1$ and the system's sensitivity to new decisions, $\frac{n_0}{k}$. What happens when we change the initial risk level, $p_0$, among defendants, and allow the rest of the process to remain the same?

Let's start by looking at what the expected value of our risk level, $p_i$, will be for time-step $i$, assuming only the prior risk $p_{i-1}$. We have from equation 5.2 that:

$$\tilde{p}_{i+1} = \tilde{p}_i \frac{n_{i-1}}{n_{i-1} + k} + \tilde{X}_i \frac{k}{n_{i-1} + k}$$

Taking the expectation over the linear equation:

$$E(p_{i+1}) = \frac{n_{i-1}}{n_{i-1} + k} E(p_i) + \frac{k}{n_{i-1} + k} E(X_i)$$

Using our knowledge that an indicator variable has expectation equal to its probability of being 1, we know:

$$E(p_{i+1}) = \frac{n_{i-1}}{n_{i-1} + k} p_i + \frac{k}{n_{i-1} + k} p_i = \frac{n_{i-1} + k}{n_{i-1} + k} p_i = p_i$$

Therefore, for any $p_i \in [0, 1]$, the urn process maintains the same expected risk level, no matter how convergent or divergent the risk becomes over sequential deci-sions. This means that if black individuals are, on average, more likely to be labeled as high-risk individuals, our model of algorithmic risk assessments will not rectify these inequalities over time.

Some, including Kleinberg, believe that algorithmic risk assessment can lower the number of black people incarcerated (Kleinberg et al., 2017). Note that this is different from rectifying *inequalities* that exist in assessments: as long as the rate of white defendants decreases by the same rate proportion, the system is still treating more black people as high-risk than whites.

However, it is important to note that varying the initial probability of conviction does not lead to divergent effects for white and black people. The static expected risk for both groups implies that an initial bias will not perpetuate or magnify biases over time, according to our model. Purportedly unbiased algorithms can perpetuate and codify existing biases, therefore, but are unlikely to lead to divergent treatment as the result of initial conditions, according to our model.

## Entrenched Algorithmic Bias

Say, instead of assuming different initial probabilities of high-risk classifications for white and black folks, we instead assume that the algorithm itself produces biased judgments each time it makes a decision. Since no algorithm in use takes in race as an explicit variable, we may assume that race is reconstructed using correlated variables. Before, our urn model looked at risk assessments as a weighted average of prior risk belief and a random variable representing the most recent risk-assessment result. Now, let's add a race indicator to our weighting system. Now, each decision is a function of prior risk, the outcome of the most recent assessment, and the race of the defendant. If we denote the race of the defendant as a variable R, and write

simply:

$$R \in \begin{cases} 1, & \text{if defendant is black} \\ 0, & \text{if defendant is white} \end{cases}$$

Then we can write the biased risk level at decision $i$ as $p_i^b$, defined below:

$$p_i^b = p_{i-1}^b [\gamma_i] + R[\rho] + X_{i-1}^b [1 - \gamma_i - \rho], \quad i \in \{2, ..., N\}, \quad \gamma_i \in [0, 1], \quad \rho \in [0, 1 - \gamma_i]$$

We don't assume $\rho$ to depend on $i$, as we might assume $\rho$ to be a function of static features that do not change over time - education level, age at first arrest, family criminal history, etc.

When this is the case, we see that the bias affects every step in the algorithm and our system converges almost surely to 1 for black people and 0 for whites, so long as $\rho > 0$. Below are simulated risk assessments for adding a weight of 0.01 to each assessment - a level of bias that could go undetected in statistical validation experiments.

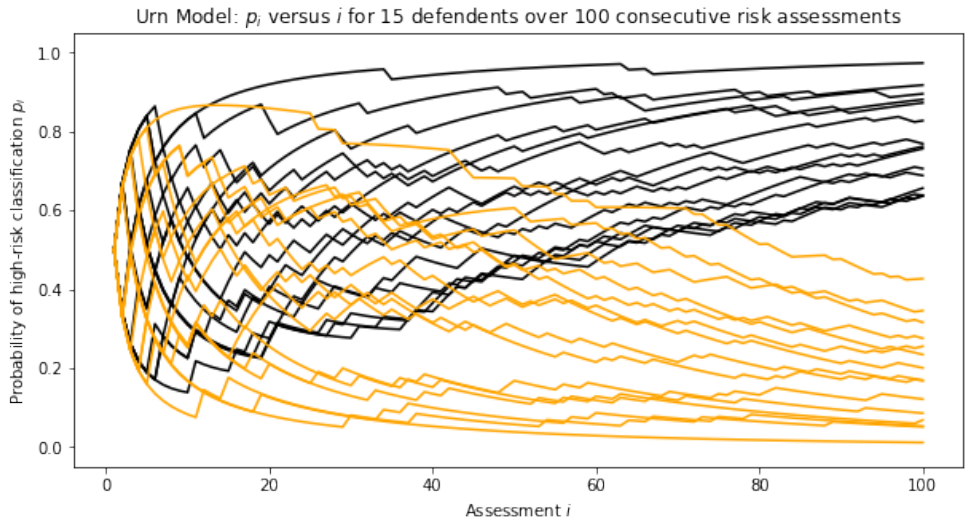

# Conclusion

After Federal oversight and military presence began to withdraw from the South at the end of Reconstruction, Southern governments promptly moved to limit black voting. Even as lynching and direct intimidation became less common, one practice - poll literacy tests - persisted in the South until the civil rights legislation of the 1960s (Goldman, 2004). Citing voter intelligence - rather than race explicitly - Southern states defended these discretionary tests against accusations of racism. The Supreme Court even upheld the tests in Lassiter vs. Northampton Election Board (1959), finding that literacy could be an appropriate basis upon which to restrict voting. In the Court's opinion, Justice William Douglas wrote:

> Residence requirements, age, previous criminal record are obvious examples indicating factors which a State may take into consideration in determining the qualifications of voters. The ability to read and write likewise has some relation to standards designed to promote intelligent use of the ballot. (*Lassiter vs. Northampton Election Board (1959), Court Opinion, 51*)

These tests have become emblematic of the historical coordinated efforts to disenfranchise black Americans. And they worked - in 1960, 30% of black southerners voted. In Mississippi, just 6.7% of black people voted, down from 70% in 1867 (Shapiro, 1993, 537-8), (Goldman, 2004, 617).

The actuarial age in criminal justice holds analogues to poll tests for three reasons. First, both purport to be race-blind, but rely on measures that resoundingly exhibit racial inequity. Even if poll tests measured literacy levels objectively and without bias, their use could reliably exclude black Americans disproportionately, because segregated education in the United States left a gap in literacy. This reality became more officially recognized after a shift in opinion from the Supreme Court. In Gaston County v. United States (1965), the Court ruled that "the County deprived its black residents of equal educational opportunities, which in turn deprived them of an equal chance to pass the literacy test" (*Gaston County v. United States, Court Opinion, 291*). Goldman in 2004 makes the analogy between poll tests' reliance on disparities in education and felony disenfranchisement relying on inequality in conviction. But institutional, discretionary reliance is not unique to felon disenfranchisement. Instead, it plagues virtually every decision in criminal punishment - arrest, bail, trial, sentencing, probation and parole. And risk assessments don't just rely on criminal history; they, too, often rely on education level, psychiatric labels, employment status, and housing.

Second, poll literacy tests and risk assessments are similar because they use self-referential logic to formalize and entrench existing power dynamics. When Southern lawmakers came under political pressure from illiterate whites whose voting rights were in jeopardy, many Southern states adopted 'Grandfather Clauses.' These clauses permitted the descendants of anybody who could vote before the Civil War to skip literacy tests. These exemptions blatantly targeted the descendants of slaves for the literacy requirement, which became nearly impossible to pass. As risk assessments moved away from using race explicitly in the 1960s, they underwent a similar trajectory: they began asking about the criminal history of family members, neighbors and peers. A released COMPAS questionnaire has a whole section devoted to 'Family Criminality' - in it, defendants must indicate if they grew up in foster care, if a parent was ever arrested or sent to jail or prison, and if a parent had a drug problem.[8] Like grandfather clauses, these questions are grasping at historical designators to justify the status quo. Proponents act like these designators are objective and 'evidence-based', and ignore their own role in persistent inequity.

Third, the two tests are similar because they're designed for people to fail. The risk principle is premised on the assumption that people should be incapacitated before they commit heinous crimes. It seeks to identify high-risk people and lock them up early. Where poll literacy tests asked questions that were effectively impossible to answer, risk assessment algorithms direct harsh treatment to people whose actions do not themselves warrant harsh punishment. By catering interventions to a defendant's risk rather than directly to behavior, the theory of incapacitation strips people's rights before they've had the chance to prove anybody wrong.

The actuarial impulse in criminal punishment - intriguing as it may be - poses a challenge to our basic commitment to equality under the law. What does this mean for criminal decisions moving forward?

Resisting prediction in criminal treatment does not mean throwing evidence out the window. I hope that this paper has exhibited one of the numerous ways that evidence (and yes, even risk assessment algorithms) can serve a purpose in analyzing criminal policy decisions and identifying biases. It can even be important for use in certain cases; for example, increased security at a crowded event might be warranted because of an anticipation of violence. But using historical, imperfect indicators to label people with risk levels, and crafting individualized punishment based on those designations, can undermine two basic goals of criminal justice - reducing violence and treating people as equals before the law.

---

[8] See COMPAS risk assessment example questions, found at https://www.documentcloud.org/documents/2702103-Sample-Risk-Assessment-COMPAS-CORE.html.